%% file: ms.tex
\pdfoutput=1
\documentclass[journal]{IEEEtran}
\makeatletter
\def\endthebibliography{%
	\def\@noitemerr{\@latex@warning{Empty `thebibliography' environment}}%
	\endlist
}
\makeatother

\PassOptionsToPackage{usenames,dvipsnames}{xcolor}

\usepackage{hyperref}

\usepackage{xparse}
\ExplSyntaxOn
\NewDocumentCommand{\storedata}{mm}{\tl_set:Nn#1{#2}}
\DeclareExpandableDocumentCommand{\getdata}{O{1}m}{\tl_item:Nn#2{#1}}
\ExplSyntaxOff

\usepackage{filecontents}
\usepackage{datatool}

\usepackage{varwidth} 
\usepackage{pdflscape}

\usepackage{fancyhdr}

\fancypagestyle{firstpage}{
  \fancyhf{}
	\fancyfoot[C]{UNCLASSIFIED: Distribution Statement A. Approved for public release; distribution is unlimited. \#28420}

}
\usepackage{amsthm}
\theoremstyle{definition}
\newtheorem{definition}{Definition:}[section]

\theoremstyle{remark}

\usepackage{amsmath}
\usepackage{cases}
\usepackage{mathrsfs}
\usepackage{amssymb}
\DeclareMathAlphabet\mathbfcal{OMS}{cmsy}{b}{n}
\usepackage{siunitx}

\usepackage{stackengine,graphicx,scalerel}
\newcommand\CircArrowRight[1]{\stackengine{-.3ex}{\scalebox{.8}{#1}}{\CAR}{O}{c}{F}{F}{L}}
\newcommand\CAR{\scaleto{\circlearrowright}{3ex}}

\usepackage[T1]{fontenc}
\usepackage{beramono}
\usepackage{listings}

\lstdefinelanguage{Julia}
  {morekeywords={abstract,break,case,catch,const,continue,do,else,elseif,
      end,export,false,for,function,immutable,import,importall,if,in,
      macro,module,otherwise,quote,return,switch,true,try,type,typealias,
      using,while},
   sensitive=true,
   alsoother={$},
   morecomment=[l]\#,
   morecomment=[n]{\#=}{=\#},
   morestring=[s]{"}{"},
   morestring=[m]{'}{'},
}[keywords,comments,strings]%

\usepackage{tikz,pgfplots,tkz-euclide}
\usetikzlibrary{shapes,arrows.meta,fit,mindmap,trees,calc,quotes,angles}

\usepackage[]{graphicx}
\usepackage{epstopdf}
\pgfplotsset{compat=newest}
\usetikzlibrary{matrix}
\usepgfplotslibrary{groupplots,colorbrewer} 
\usetikzlibrary{pgfplots.colorbrewer,datavisualization}

\definecolor{s1}{RGB}{228, 26, 28}
\definecolor{s2}{RGB}{55, 126, 184}
\definecolor{s3}{RGB}{77, 175, 74}
\definecolor{s4}{RGB}{152, 78, 163}
\definecolor{s5}{RGB}{255, 127, 0}

\pgfplotscreateplotcyclelist{my markers}{%
{black, solid, every mark/.append style={solid, fill=black},mark = *},
{s1, dotted, every mark/.append style={solid, fill=s1},mark = halfsquare left*},
{s2, densely dashdotted, every mark/.append style={solid, fill=s2},mark = triangle*},
{s3, loosely dotted, every mark/.append style={solid, fill=s3, scale = 1},mark = diamond*},
{s4, dashed, every mark/.append style={solid, fill=s4, scale = 0.5}, mark = otimes*},  
{s5, densely dotted, every mark/.append style={solid, fill=s5},mark = halfcircle*},
{s1,dashdotted, every mark/.append style={solid, fill=s1},mark = 10-pointed star},
{s2,densely dashdotted, every mark/.append style={solid, fill=s2},mark = pentagon},
{s3,solid, every mark/.append style={solid},mark = ball,ball color=s3},
}

\pgfplotscreateplotcyclelist{my markersB}{%
{s2, densely dashdotted, every mark/.append style={solid, fill=s2},mark = triangle*},
{s3, loosely dotted, every mark/.append style={solid, fill=s3, scale = 1},mark = diamond*},
{s4, dashed, every mark/.append style={solid, fill=s4, scale = 0.5}, mark = otimes*},  
{s5, densely dotted, every mark/.append style={solid, fill=s5},mark = halfcircle*},
{s1,dashdotted, every mark/.append style={solid, fill=s1},mark = 10-pointed star},
{s2,densely dashdotted, every mark/.append style={solid, fill=s2},mark = pentagon},
{s3,solid, every mark/.append style={solid},mark = ball,ball color=s3},
{black, dashed, every mark/.append style={solid, fill=s4, scale = 1}, mark = otimes*},
{s2, loosely dotted, every mark/.append style={solid, fill=s2, scale = 1},mark = diamond*},
}

\pgfplotscreateplotcyclelist{my markersC}{%
{s2, densely dashdotted, every mark/.append style={solid, fill=s2},mark = triangle*},
}

\pgfplotscreateplotcyclelist{my markersA}{%
{black, solid, every mark/.append style={solid, fill=black},mark = *},
{s1, dotted, every mark/.append style={solid, fill=s1, scale = 0.5},mark = halfsquare left*},
{s2, densely dashdotted, every mark/.append style={solid, fill=s2},mark = triangle*},
{s3, loosely dotted, every mark/.append style={solid, fill=s3, scale = 1},mark = diamond*},
{s4, dashed, every mark/.append style={solid, fill=s4, scale = 0.5}, mark = otimes*},  
{s5, densely dotted, every mark/.append style={solid, fill=s5},mark = halfcircle*},
{s1,dashdotted, every mark/.append style={solid, fill=s1},mark = 10-pointed star},
{s2,densely dashdotted, every mark/.append style={solid, fill=s2},mark = pentagon},
{s3,solid, every mark/.append style={solid},mark = ball,ball color=s3},
}

\pgfplotscreateplotcyclelist{my markers ee}{%
{black, solid, every mark/.append style={solid, scale =.7},mark = otimes*},
{s2, dotted, every mark/.append style={solid, scale =.4},mark = triangle*},
{densely dashdotted, every mark/.append style={solid, fill=magenta},mark = triangle*},
{loosely dotted, every mark/.append style={solid, fill=violet, scale = 1},mark = diamond*},
{teal, dashed, every mark/.append style={solid, fill=purple}, mark = otimes*},  
{red, densely dotted, every mark/.append style={solid, fill=green!60!black},mark = halfcircle*},
{dashdotted, every mark/.append style={solid, fill=green},mark = 10-pointed star},
{densely dashdotted, every mark/.append style={solid, fill=black},mark = pentagon},
{solid, every mark/.append style={solid, fill=red},mark = ball,ball color=green},
}

\pgfplotsset{every axis plot/.append style= ultra thick }
\pgfplotscreateplotcyclelist{my black white}{%
solid, every mark/.append style={solid, fill=gray}, mark=*\\%
dotted, every mark/.append style={solid, fill=gray}, mark=square*\\%
densely dotted, every mark/.append style={solid, fill=gray}, mark=otimes*\\%
loosely dotted, every mark/.append style={solid, fill=gray}, mark=triangle*\\%
dashed, every mark/.append style={solid, fill=gray},mark=diamond*\\%
loosely dashed, every mark/.append style={solid, fill=gray},mark=*\\%
densely dashed, every mark/.append style={solid, fill=gray},mark=square*\\%
dashdotted, every mark/.append style={solid, fill=gray},mark=otimes*\\%
solid, every mark/.append style={solid},mark=star\\%
densely dashdotted,every mark/.append style={solid, fill=gray},mark=diamond*\\%
}
\usepackage{pgfplotstable}
\pgfplotstableset{
  col sep=comma,
string replace*={_}{\textsubscript},
every head row/.style={before row=\toprule,after row=\midrule},
every last row/.style={after row=\bottomrule},
display columns/0/.style={string type,column name={}},
empty cells with={NaN} 
}

\usepackage{svg}
\usepackage{lineno,hyperref}
\modulolinenumbers[5]
\definecolor{links}{HTML}{2A1B81}
\hypersetup{colorlinks,linkcolor=,urlcolor=links}

\tikzstyle{block} = [draw,rectangle,thick,minimum height=2em,minimum width=2em,align=center]
\tikzstyle{sum} = [draw,circle,inner sep=0mm,minimum size=2mm]
\tikzstyle{connector} = [->,thick]
\tikzstyle{output} = [coordinate]

\usepackage{outlines}
\colorlet{ColorPink}{red!10}

\newcommand{\NLOpt}{\texttt{NLOptControl} }
\newcommand{\NLOptnsp}{\texttt{NLOptControl}}

\newcommand{\R}{\mathbb{R}}
\newcommand{\C}{\mathbb{C}}

\newcommand{\TaxName}[1]{\emph{#1}}

\newcommand{\GPOPSiinsp}{\TaxName{GPOPS-ii}}

\newcommand{\GPOCSnsp}{\TaxName{GPOCS}}

\newcommand{\MATLAB}{\TaxName{MATLAB} }

\newcommand{\IPOPT}{\TaxName{IPOPT} }
\newcommand{\KNITRO}{\TaxName{KNITRO} }

\storedata\NLstringsA{{Solution}{Collocation Pts.}}
\storedata\NLstringsC{{\NLOpt}{\texttt{NLOpt.} Colloc. Pts.}}
\storedata\PRstringsA{{\PROPT}{\PROPT Colloc. Pts.}}
\storedata\NLstringsB{{nl1}{nl2}{nl4}{nle}{nlt}}
\storedata\PRstringsB{{pr1}{pr2}{pr4}}

\usepackage{caption}

\def\slides{0}
\def\thesis{0}

\newcommand{\reqA}{$S_1$ }
\newcommand{\reqAnsp}{$S_1$}
\newcommand{\reqB}{$S_2$ }
\newcommand{\reqBnsp}{$S_2$}
\newcommand{\reqC}{$S_3$ }
\newcommand{\reqCnsp}{$S_3$}
\newcommand{\reqD}{$S_4$ }
\newcommand{\reqDnsp}{$S_4$}
\newcommand{\reqE}{$S_5$ }
\newcommand{\reqEnsp}{$S_5$}
\newcommand{\reqF}{$S_6$ }
\newcommand{\reqFnsp}{$S_6$}
\newcommand{\reqG}{$S_7$ }
\newcommand{\reqGnsp}{$S_7$}

\newcommand{\plannerA}{$P_A$ }
\newcommand{\plannerAnsp}{$P_A$}
\newcommand{\plannerB}{$P_B$ }
\newcommand{\plannerBnsp}{$P_B$}
\newcommand{\plannerC}{$P_C$ }
\newcommand{\plannerCnsp}{$P_C$}
\newcommand{\plannerD}{$P_D$ }
\newcommand{\plannerDnsp}{$P_D$}

\newcommand{\caseA}{$E_A$ }
\newcommand{\caseAnsp}{$E_A$}
\newcommand{\caseB}{$E_B$ }
\newcommand{\caseBnsp}{$E_B$}
\newcommand{\caseC}{$E_C$ }
\newcommand{\caseCnsp}{$E_C$}

\begin{document}

\title{Real-time trajectory planning for automated vehicle safety and performance in dynamic environments}
\author{Huckleberry~Febbo, Paramsothy~Jayakumar, Jeffrey~L.~Stein, and Tulga~Ersal$^{*}$%
\thanks{The authors wish to acknowledge the financial support of the Automotive Research Center (ARC) in accordance with Cooperative Agreement W56HZV-14-2-0001 U.S. Army Tank Automotive Research, Development and Engineering Center (TARDEC) Warren, MI.,DISTRIBUTION A. Approved for public release; distribution unlimited.}%
\thanks{H. Febbo, J. L. Stein, and T. Ersal are with the Department of Mechanical Engineering, University of Michigan, Ann Arbor, MI 48109 USA (e-mail: febbo@umich.edu; stein@umich.edu; tersal@umich.edu).}%
\thanks{P. Jayakumar is with the U.S. Army RDECOM-TARDEC, Warren, MI 48397 (email: paramsothy.jayakumar.civ@mail.mil)}%
\thanks{$^{*}$Corresponding author (tersal@umich.edu)}}
\maketitle

\begin{abstract}
Safe trajectory planning for high-performance automated vehicles in an environment with both static and moving obstacles is a challenging problem. Part of the challenge is developing a formulation that can be solved in real-time while including the following set of specifications: minimum time-to-goal, a dynamic vehicle model, minimum control effort, both static and moving obstacle avoidance, simultaneous optimization of speed and steering, and a short execution horizon. This paper presents a nonlinear model predictive control-based trajectory planning formulation, tailored for a large, high-speed unmanned ground vehicle, that includes the above set of specifications. This paper also evaluates \NLOptnsp's ability to solve this formulation in real-time in conjunction with the \KNITRO nonlinear programming problem solver; \NLOpt is our open-source, direct-collocation based, optimal control problem solver. This formulation is tested with various sets of the specifications. In particular, a parametric study relating execution horizon and obstacle speed, indicates that the moving obstacle avoidance specification is not needed for safety when the planner has a small execution horizon ($\leq0.375\;\si{s}$) and the obstacles are moving slowly ($\leq2.11\frac{\si{m}}{\si{s}}$). However, a moving obstacle avoidance specification is needed when the obstacles are moving faster, and this specification improves the overall safety by a factor of $6.73$ ($p=2.2\times10^{-16}$) without, in most cases, increasing the solve-times. Overall, the results indicate that (1) safe trajectory planners for high-performance automated vehicles should include the entire set of specifications mentioned above, unless a static or low-speed environment permits a less comprehensive planner; and (2) \NLOpt can solve the formulation in real-time.
\end{abstract}

\section{Introduction}
In many high-performance automated vehicle applications, e.g., in unmanned air vehicles (UAVs), unmanned ground vehicles (UGVs), and spacecraft, it is both desirable and challenging to plan safe trajectories in a dynamic environment. Part of this challenge is incorporating the set of specifications listed in Table \ref{tab:cap} into a real-time planner, where real-time planning demands that the planner's solve-times are all less than the execution horizon. While trajectory planning systems that include subsets of the specifications listed in Table \ref{tab:cap} exist, a planner that consists of all of them has not yet been developed.

\begin{figure}[!t]
\begin{center}
\captionof{table}{Planner Specifications} \label{tab:cap}
\begin{tabular}{ l l}
\hline
specification & Description \\
\hline
\reqA & static obstacle avoidance \\
\reqB & minimum time-to-goal \\
\reqC & dynamic vehicle model \\
\reqD & minimum control effort \\
\reqE & simultaneously optimize speed and steering \\
\reqF & moving obstacle avoidance \\
\reqG & small execution horizon \\
\end{tabular}
\end{center}
\end{figure}

Fig. \ref{fig:terms} shows a conceptual scheme for comparing and developing trajectory planners. This scheme illustrates the conceptual performance and safety of a vehicle controlled using trajectory planners with different sets of specifications, operating either in a static environment with a stationary obstacle (top four traces) or a dynamic environment with a moving obstacle (bottom two traces). In all cases, the planning and execution horizons are the same.

\if\thesis1
\newcommand{\figAW}{0.7}
\else
\newcommand{\figAW}{0.95}
\fi

\begin{figure}
\begin{minipage}[c]{\linewidth}
 \begin{center}
  \includegraphics[width=\figAW\textwidth]{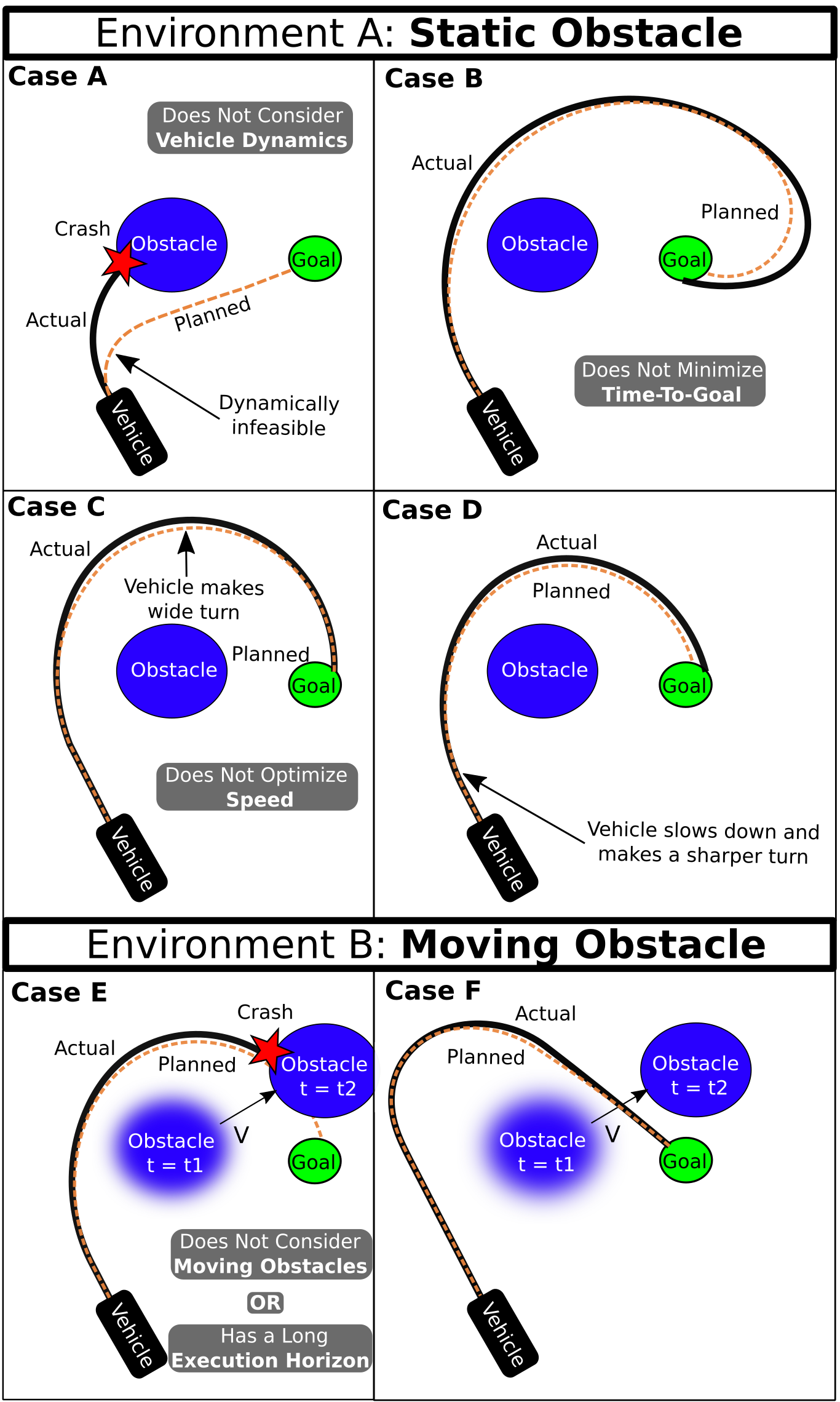}  \caption{Comparison of trajectory planners illustrating the conceptual effect that planner specifications have on performance and safety within a given environment. \label{fig:terms}}
  \end{center}
\end{minipage}
\end{figure}

Static obstacle avoidance (\reqAnsp, Table \ref{tab:cap}) is a baseline specification in many trajectory planning systems, but is not, by itself, sufficient for either performance or safety. Regarding safety, if the trajectory planner does not use a dynamic vehicle model (\reqCnsp, Table \ref{tab:cap}), a trajectory that the vehicle cannot follow may be determined \cite{frasch2013auto}. Such a trajectory may result in either a collision \cite{frasch2013auto,frazzoli2002real} (Case A, Fig. \ref{fig:terms}) or some other catastrophic event, such as rollover in the case of a ground vehicle \cite{lo2017high}. Despite this, some planners designed to avoid static obstacles for UAV applications \cite{greatwood2013} utilize a kinematic vehicle model (Case A, Fig. \ref{fig:terms}). By utilizing a dynamic vehicle model in trajectory planning, the actual vehicle can follow the prescribed trajectory more accurately. Planners designed to avoid static obstacles with a dynamic vehicle model (Case B, Fig. \ref{fig:terms}) exist for UGV applications \cite{yoon2009model}. However, the planner in Case B does not have a minimum time-to-goal specification (\reqBnsp, Table \ref{tab:cap}), which may result in failure for certain applications. For instance, in racing applications \cite{Gerdes_2012,velenis2007aggressive}, planning without this specification will likely result in a lost race. In these applications, the planner should include at least \reqAnsp-\reqC (Case C, Fig. \ref{fig:terms}), such that it can arrive at the goal in less time than a planner with only the static obstacle avoidance and dynamic vehicle model specifications. If minimizing fuel consumption and mechanical wear are additional concerns, then the minimum control effort specification (\reqDnsp, Table \ref{tab:cap}) needs to be included in the planner as well. Planners with \reqAnsp-\reqD exist in applications for UGVs \cite{liu2016study} and UAVs \cite{liu2017search,mohtaexperiments, liu2017search2}. A limitation of these planners is that they do not optimize both speed and steering (\reqEnsp, Table \ref{tab:cap}). Optimizing both allows the vehicle to both slow down more quickly and turn more tightly (shown in Case D, Fig. \ref{fig:terms}), thereby improving both performance and safety \cite{liu2017combined}.

In a dynamic environment, while the trajectory planning specifications \reqAnsp-\reqE are necessary for both performance and safety, they are not sufficient (see Case E, Fig. \ref{fig:terms}). To improve collision avoidance (i.e., safety) in a dynamic environment there are three possibilities: including a moving obstacle avoidance specification (\reqFnsp, Table \ref{tab:cap}); including a small execution horizon specification (\reqGnsp, Table \ref{tab:cap}); or including both.

A moving obstacle avoidance specification accounts for the motion of the obstacle over the planning horizon, which increases safety (see Case F, Fig. \ref{fig:terms}). This specification has been implemented for applications in UGVs \cite{nishio2017moving}, UAVs \cite{yao2015real}, and spacecraft \cite{jewison2015model}. These developments, however, have a limitation: they use a kinematic vehicle model as opposed to a dynamic vehicle model; Case A, Fig. \ref{fig:terms} depicts the potential outcome of using a kinematic vehicle model. Our preliminary work \cite{febbo2017moving} developed a planner with \reqAnsp-\reqF for a UGV application. This work, however, has several limitations, three of which are: it does not investigate closed-loop performance and safety; it assumes that the goal is within the LiDAR's sensing range; and, finally, the planner's solve-times are, at best, nearly two orders of magnitude above real-time (assuming an execution horizon of $0.5\;\si{s}$). Among other things, this paper addresses these three limitations.

A small execution horizon\footnote{An execution horizon is described as "small" when reducing it does not improve safety within a given environment.} specification engenders a more reactive planner with better obstacle avoidance capabilities. For instance, to avoid the collision in Case E (Fig. \ref{fig:terms}), a smaller execution horizon can be used. Previous research \cite{liu2017combined} includes a small execution horizon as well as \reqAnsp-\reqEnsp. While there is reason to expect that such a planner may operate safely around slowly moving obstacles, this hypothesis has not yet been tested. Therefore, this paper also investigates, for the first time, whether a system with \reqAnsp-\reqE and a small execution horizon can operate safely in a dynamic environment for a range of obstacle speeds.

The ultimate goal of this research is to develop a trajectory planning formulation that has all of the specifications listed in Table \ref{tab:cap}. The motivation for this investigation is the assumption that a planner with this set of specifications would represent an improvement in both safety and performance over planners with less comprehensive sets of specifications.

This work uses a nonlinear model predictive control (NMPC)-based trajectory planner; this approach is also used in \cite{greatwood2013,yoon2009model,liu2017search,mohtaexperiments,liu2017search2, geisert2016trajectory,liu2017combined,nishio2017moving,yao2015real,jewison2015model}. Unfortunately, it is very difficult to solve the proposed planning formulation in real-time with a short execution horizon. For instance, the literature shows that \GPOCSnsp, \GPOPSiinsp, and our custom software, all written in the MATLAB computation language, are not fast enough for NMPC applications in aircraft \cite{basset2010computing}, robot \cite{gpops_robotA}, and UGV \cite{liu2017combined,febbo2017moving} systems, respectively. As part of this work, \NLOptnsp's ability to solve the proposed formulation in real-time wi short execution horizon is tested; \NLOpt is our open-source, direct-collocation based optimal control problem (OCP) solver \cite{febbo_2017}. As an example, the trajectory planning formulation developed in this work is tailored for a high mobility multipurpose wheeled vehicle (HMMWV), but can be adapted to other vehicles as well. Together, this planner and vehicle are referred to as an UGV.

This paper addresses the following research objectives:
\begin{itemize}
 \item Introduce an NMPC-based trajectory planner with \reqAnsp-\reqGnsp, tailored for a UGV application.
 \item Investigate the effect that different sets of specifications have on safety, performance, and solve-time. 
 \item Investigate the need to include a moving obstacle avoidance specification for a range of execution horizons and obstacle speeds.
 \item Investigate \NLOptnsp's ability to solve the proposed formulation in real-time with a short execution horizon.
\end{itemize}

This paper assumes that
\begin{itemize}
\item both the goal and obstacle information are known,
\item the vehicle state is known, and
\item the terrain is flat.
\end{itemize}

The remainder of this paper is organized as follows: Section~\ref{sec:math} describes the NMPC framework developed to consider non-negligible trajectory planning problem solve-times and the underlying OCP formulation developed to include \reqAnsp-\reqFnsp. Section~\ref{sec:methods} describes the test conditions under which the proposed planner is evaluated. In Section~\ref{sec:resultsB}, the effect that adding different specifications to trajectory planners has on safety, performance, and solve-time is tested in a variety of environments. The results of these tests are discussed in Section~\ref{sec:discussion}. Section~\ref{sec:conclusion} summarizes the paper and draws conclusions.
\section{Mathematical Formulation} \label{sec:math}
\subsection{NMPC Framework}
\begin{figure}
\begin{center}
\begin{tikzpicture}[scale=1, auto, >=stealth']
\small
\lineskip -2pt
 \matrix[ampersand replacement=\&, row sep=1.15cm, column sep=1.32cm]{
 \node [block, name=OCP] {OCP}; \& \node [output] (u1) {}; \& \node [block, name=V] {Vehicle \\ Model};  \\
 \node [output] (q1) {}; \& \node [block, name=E] { State \\ Prediction };  \& \node [output] (q2) {}; \\
                                                                         };
\node [output, left of=OCP] (IN) {};
\node [output, below of=E] (ui) {};
\huge
\node [above of=E, label={[shift={(-1.5,-0.9)}]\CircArrowRight{$t_{ex}$}}] {};
\small
\node [right of=OCP, label={[shift={(0,0)}]$\boldsymbol{\zeta}^*$}] {};
\node [output, right of=E] (RE) {};
\node [output, left of=E] (LE) {};
\node [output, above of=V] (iii) {};

    \node [output, below of=OCP] (A) {};
    \node [output, below of=V] (B) {};
     \draw [connector] ([xshift=0em]ui) -- node[name=uo] {$\mathbf{U0}$} ([xshift=0em]E);
    \draw [connector] ([xshift=-2em]IN.west) -- node[name=L] {$\mathbfcal{G},\mathbfcal{E}$} (OCP);
    \draw [connector] (iii) -- node {$\mathbf{X0},\mathbf{U0}$} (V);
    \draw [connector] (OCP) -- (V);
    \draw [connector] (V) -- (q2) -- node[name=XA] {$\mathbf{X0}$} (E);
    \draw [connector] (E) -- (q1) -| node[name=XB] {$\mathbf{X0}_p$} (OCP);
    \draw [connector] ([xshift=0cm]u1) -- node[name=UB] {}  ([xshift=0cm]E);
\end{tikzpicture}
\caption{Nonlinear model predictive control framework used to account for non-negligible optimal control problem (OCP) solve-times.}\label{fig:mpc}
\end{center}
\end{figure}
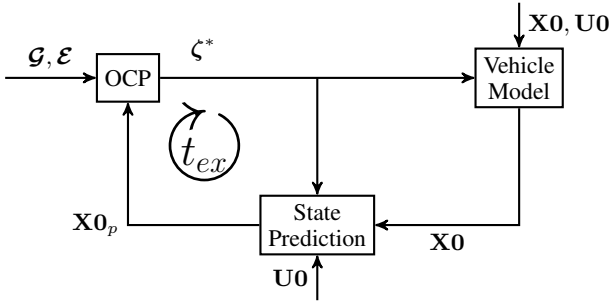

At heart of an NMPC formulation lies an OCP. In NMPC simulation studies, OCP solve-times are often neglected \cite{simon2009swelling,liu2017combined}. In such a case, first the plant simulation is paused, and the OCP is initialized at the current time $t_0$ with the current plant state $\mathbf{X0}$. Next, the OCP is solved to produce an optimal control signal $\boldsymbol{\zeta}^*$ (i.e., trajectory). With this signal, the plant is then simulated starting at $t_0$ with $\mathbf{X0}$ until $t_0+t_{ex}$; this yields a new initial state, which is then used to initialize the next OCP. However, most practical OCPs take a non-negligible amount of time to solve, after which, in a more realistic simulation, the plant will have evolved from its current state, where the OCP was initialized, to a new state \cite{simon2009swelling}. This computational delay renders the control signal sub-optimal and potentially infeasible or unsafe. To achieve optimal safety and performance, non-negligible OCP solve-times must be taken into account. The NMPC framework used in this work, shown in \ref{fig:mpc}, accounts for these non-negligible OCP solve-times.

This framework has three main components: the OCP, the vehicle model (or plant model), and the state prediction function. The OCP is provided with goal and environment information, defined as follows:
\begin{definition}{Goal information $\mathbfcal{G}$}
includes the goal position ($x_g,y_g$), the desired vehicle orientation at the goal $\psi_g$, and the radial tolerance for attaining the goal $\sigma$.
\end{definition}
\begin{definition}{Environment information $\mathbfcal{E}$}
includes the sizes, initial positions, and velocities of the obstacles.
\end{definition}
The obstacles are assumed to be ellipse-shaped, where $\mathbf{a_{obs}}$ and $\mathbf{b_{obs}}$ are arrays that describe obstacles' semi-major and semi-minor axes, respectively; $\mathbf{x0_{obs}}$ and $\mathbf{y0_{obs}}$ are arrays of the obstacles' initial $x$ and $y$ positions, respectively; and $\mathbf{v_x}$ and $\mathbf{v_y}$ are arrays of the obstacles' speeds in the $x$ and $y$ directions, respectively.

During the first execution horizon, the OCP has not produced a control signal for the vehicle to follow. Therefore, the vehicle is sent a known control signal $\mathbf{U0}$ (see Fig. \ref{fig:mpc}), set such that the vehicle will drive straight at a constant speed. To account for the evolution of the plant state during the execution horizon, a state prediction $\mathbf{X0}_p$, that is made for $t_0 + t_{ex}$, is used to initialize the OCP. The inputs of the state prediction function are the current state of the vehicle and the current control signal, which is $\mathbf{U0}$ during the first execution horizon and $\boldsymbol{\zeta}^*$ afterward. Then, the plant model is simulated from the initial time $t_0$ to $t_0 + t_{ex}$ and the first OCP is solved. Real-time feasibility of this framework requires that the OCP solve-times be all less than the execution horizon.

Model mismatch between the plant model, the vehicle dynamics model in the OCP, and the state prediction function can induce biases. These biases can affect the integrity of the research objectives, provided in the previous section. Addressing these biases is acknowledged as an important research problem, but is not one of the goals of this paper. Therefore, to avoid these biases while focusing on our research tasks, in this paper, the plant model, the vehicle model in the OCP, and the state prediction function all use the same set of differential equations, which is presented in detail later in this section.

This framework runs until the UGV either reaches the goal or fails the test. An algorithm is run after each execution horizon to determine if the vehicle has reached the goal within the radial goal tolerance $\sigma$. The test fails if
\begin{enumerate}
  \item the vehicle crashes into an obstacle,
  \item the vertical tire load in the plant model goes below $100\;\si{N}$ for any of the four tires,
  \item any of the solve-times exceeds $300\;\si{s}$, or if
  \item the solution to the nonlinear programming problem (NLP) is not considered to be optimal based on the tolerances and the Karush-Kuhn-Tucker conditions.
\end{enumerate}

\subsection{Optimal Control Problem}\label{sec:ocp}
This section describes how the set of planner specifications \reqAnsp-\reqF are incorporated into the OCP. At a high-level, these specifications are all incorporated into the single-phase, continuous-time OCP defined in Eqn. \ref{eqn:cost} - Eqn. \ref{eqn:event} as
\begin{align}
   & \underset{\xi(t),\;\zeta(t),\;t_f}{\textbf{minimize}} & {\cal{M}}(\xi(t_0+t_{ex}),t_0+t_{ex},\xi(t_f),t_f) + \nonumber & & \\
   & & \int_{t_0+t_{ex}}^{t_f}L(\xi(t),\zeta(t),t) \, dt \label{eqn:cost} & &\\
   & \textbf{subject to} & & & \nonumber \\
   & & \frac{d\xi}{dt}(t) - f(\xi(t),\zeta(t),t) =0\label{eqn:dynamics} & &\\
   & & C(\xi(t),\zeta(t),\mathcal{A}(t),t_f) \le 0\label{eqn:path} & & \\
   & & \phi(\xi(t_0+t_{ex}),t_0+t_{ex},\xi(t_f),t_f)=0 \label{eqn:event} & &
\end{align}
where $t_f$ is the free final time, $t\in [t_0+t_{ex},t_f]$ is the time, $\xi(t) \in \R^{n_{st}}$ is the state and $\zeta(t) \in \R^{n_{ctr}}$ is the control, with $n_{st}$ defined as the number of states and $n_{ctr}$ defined as the number of controls. The Mayer term is ${\cal{M}}: \R^{n_{st}} \times \R \times \R^{n_{st}} \times \R \rightarrow \R$ and the Lagrange term is $L:\R^{n_{st}} \times \R^{n_{ctr}} \times \R \rightarrow \R$. The dynamic constraints are given by $f:\R^{n_{st}} \times \R^{n_{ctr}} \rightarrow \R^{n_{st}}$. The path constraints are captured by $C:\R^{n_{st}} \times \R^{n_{ctr}} \times \R^a \times \R \rightarrow \R^p$, which bound: the state and control based on the vehicle's dynamic limits, and any additional information, denoted as $\mathcal{A}(t) \in \R^a$; and $t_f$ based on a maximum final time $t_{f_{max}}$. Finally, the event constraints are expressed with $\phi:\R^{n_{st}} \times \R \times \R^{n_{st}} \times \R \rightarrow \R^q$.

The remainder of this section describes how \reqAnsp-\reqF are incorporated into Eqn. \ref{eqn:cost} - Eqn. \ref{eqn:event}.
\paragraph{Cost Functional}
First, the cost functional in Eqn. \ref{eqn:cost} is set to Eqn. \ref{eqn:cost_a} as
\begin{align}
&J=w_t{t_f}  \nonumber \\
& + w_g \frac{(x(t_f) - x_g)^2 + (y(t_f) - y_g)^2}{(x(t_0+t_{ex}) - x_g)^2 + (y(t_0+t_{ex}) - y_g)^2 + \epsilon} \nonumber \\
& + w_{cf} {\int_{t_0+t_{ex}}^{t_f}[ {w_{\delta_f}}{\delta_f(t)^2} +  {w_{\gamma}}{\gamma(t)^2} + w_J J_x(t)^2] \text{d} t} \nonumber \\
 & + {w_{F_z}} {\int_{t_0+t_{ex}}^{t_f}[ \text{tanh}( - \frac{F_{z_{rl}} - a}{b}) + \text{tanh}( - \frac{F_{z_{rr}} - a}{b})] \text{d} t} + \varrho  \nonumber \\
 & + w_{haf}{\int_{t_0+t_{ex}}^{t_f} [ \sin(\psi_g)(x - x_g) - \cos(\psi_g)(y - y_g)]^2 \text{d} t}
 \label{eqn:cost_a}
\end{align}
where $w_t,w_g,w_{cf},w_{\delta_f},w_{\gamma},w_J,w_{F_z},w_{haf}$ are weight terms, $x(t)$ and $y(t)$ represent the vehicle's global position coordinates on a flat plane, $\epsilon$ is a small number set to $0.01$ to avoid singularities, $\delta_f(t)$ is the steering angle at the front of the vehicle, $\gamma(t)$ is the steering rate, $J_x(t)$ is the longitudinal jerk, $a$ and $b$ are parameters to prevent $F_{z_{rl}}$ and $F_{z_{rr}}$ from being close to the minimum vertical tire load limit, denoted as $F_{z_{min}}$, as described in \cite{liu2017combined}, and $\varrho$ is a term for penalizing the slack variables on the initial and terminal conditions.
%

There are six terms in Eqn. \ref{eqn:cost_a}, the first of which minimizes the final time $t_f$, which helps establish a minimum-time-to-goal specification. The second term helps the vehicle reach the goal when the goal is not within the LiDAR range, denoted as $L_{\mathrm{range}}$. If the goal is within a distance of $L_{\mathrm{range}}$, then $w_g$ is set to zero, and the vehicle is constrained to reach the goal. This constraint is described in greater detail later in this section. The third term minimizes the control effort, which encourages smooth control signals. The fourth term dissuades the controller from generating solutions near the minimum vertical tire load limit. This is done to prevent vehicle rollover and infeasible initializations in the next OCP. The fifth term establishes soft constraints on the initial and terminal conditions. This term is also described in greater detail later in this section. Finally, to help the vehicle pass the goal location through the desired direction $\psi_g$ the sixth term is added, which minimizes the area between a line in the $(x,y)$ plane going through the goal in the desired direction $\psi_g$ and the vehicle's trajectory in the $(x,y)$ plane \cite{liu2017combined}.
\input{figs/threeDOF.tex}

\paragraph{Dynamic Vehicle Model} \label{sec:dyn_con}
When including both the dynamic vehicle model and minimum time-to-goal specifications, it is important to consider that the vehicle may need to operate at its dynamic limits. Thus, this work leverages the $3$DoF vehicle model developed in \cite{liu2016study} (shown in Fig. \ref{fig:model}). This model is designed to plan trajectories that operate a HMMWV safely at its dynamic limits \cite{liu2016study}. To achieve this, it has eight states, two controls, uses a pure-slip Pacejka tire model \cite{Pacejka_2012}, and considers the longitudinal load transfer effects when calculating the vertical tire forces. The differential equations that are used to model the plant, the vehicle dynamics in the OCP, and the state prediction function are shown in Eqn. \ref{eqn:dynamics2} as
\begin{equation}
f(\xi(t),\zeta(t),t) = \mathcal{D}(\xi(t))+\mathcal{B}\zeta(t) \label{eqn:dynamics2}
\end{equation}
where,
  \begin{align*}
    & \mathcal{D}(\xi(t)) =
     \begin{bmatrix}
        U(t)\cos\Psi(t) - (V(t) + L_f \omega_z(t))\sin\Psi(t)\\
        U(t)\sin\Psi(t) + (V(t) + L_f \omega_z(t))\cos\Psi(t)\\
        (F_{yf}(t) + F_{yr}(t))/M_t - U(t) \omega_z(t)\\
        (F_{yf}(t)L_f - F_{yr}(t)L_r)/I_{zz}\\
        \omega_z(t)\\
        0\\
        a_x(t)\\
        0
      \end{bmatrix}
          \text{,  }
  \end{align*}
  \begin{align*}
          \xi(t) =
     \begin{bmatrix}
          x(t)\\
          y(t)\\
          V(t)\\
          \omega_z(t)\\
          \Psi(t)\\
          \delta_f(t)\\
          U(t)\\
          a_x(t)
      \end{bmatrix}
  \end{align*}

  \begin{align*}
      \mathcal{B}^{T} =
      \begin{bmatrix}
        0&0&0&0&0&1&0&0\\
        0&0&0&0&0&0&0&1
        \end{bmatrix}
        \text{  and  }
        \zeta(t) =
        \begin{bmatrix}
        \gamma(t)\\
        J_x(t)\\
      \end{bmatrix}
   \end{align*}
Eqn. \ref{eqn:dynamics2} breaks the dynamics constraints in Eqn. \ref{eqn:dynamics} into two terms. The first of these terms, $\mathcal{D}(\xi(t))$, establishes the state dynamics for the global position of the vehicle, the lateral speed $V(t)$, the yaw rate $\omega_z(t)$, the heading angle $\Psi(t)$, the steering angle $\delta_f(t)$, the longitudinal speed $U(t)$, and the longitudinal acceleration $a_x(t)$. The second term, $\mathcal{B}\zeta(t)$, relates state variable rates to their respective control variables, i.e., $ \frac{d\delta_f}{dt}(t)$ to the steering rate $\gamma(t)$, and $\frac{da_x}{dt}(t)$ to the longitudinal jerk $J_x(t)$. Finally, $L_f$ and $L_r$ are the distances from the front and rear axles to center-of-mass (COM), $I_{zz}$ is the moment of inertia about the COM, $F_{yf}$ and $F_{yr}$ are the front and rear lateral tire forces, and $M_t$ is the total vehicle mass. Table \ref{tab:veh}, which is in  Appendix \ref{appendixa}, contains all of the vehicle parameters used in this paper.

The vertical tire load on each of the four tires is constrained to be above the minimum vertical tire load limit $F_{z_{min}}$. These constraints helps prevent vehicle roll-over and are incorporated into Eqn. \ref{eqn:path}. To calculate the vertical loads on the tires, this work uses a vertical load transfer model \cite{liu2017combined}. The vertical tire forces are approximated as
\begin{align*}
  F_{z_{rl}} = \frac{1}{2}(F_{z_{r0}} + K_{z_{x}}(a_x(t) - V(t)\omega_z(t) ) - K_{z_{yr}}\frac{F_{yf} + F_{yr}}{M_t} \nonumber \\
  F_{z_{rr}} = \frac{1}{2}(F_{z_{r0}} + K_{z_{x}}(a_x(t) - V(t)\omega_z(t) ) + K_{z_{yr}}\frac{F_{yf} + F_{yr}}{M_t} \nonumber \\
  F_{z_{fl}} = \frac{1}{2}(F_{z_{f0}} - K_{z_{x}}(a_x(t) - V(t)\omega_z(t) ) - K_{z_{yf}}\frac{F_{yf} + F_{yr}}{M_t} \nonumber \\
  F_{z_{fr}} = \frac{1}{2}(F_{z_{f0}} - K_{z_{x}}(a_x(t) - V(t)\omega_z(t) ) + K_{z_{yf}}\frac{F_{yf} + F_{yr}}{M_t} \nonumber \\
\end{align*}
where $F_{z_{rl}}$ and $F_{z_{rr}}$ are the rear left and rear right vertical tire loads, $F_{z_{fl}}$ and $F_{z_{fr}}$ are the front left and front right vertical tire loads, $F_{z_{r0}}=\frac{M_t L_f g}{L_f+L_r}$ is the static rear axle load, $F_{z_{f0}}=\frac{M_t L_r g}{L_f+L_r}$ is the static front axle load, and $K_{z_{x}}$ is the longitudinal load transfer coefficient, $K_{z_{yf}}$ and $K_{z_{yr}}$ are the front and rear lateral load transfer coefficients \cite{liu2017combined}.

\paragraph{State and Control Limits}
Actuator and other physical plant limits help establish the state and control bounds, which are added to Eqn. \ref{eqn:path}. Specifically, five of the states and both controls are bounded with constant upper and lower bounds as
\begin{align*}
  & x_{\min} \leq x(t) \leq x_{\max} \nonumber \\
   & y_{\min} \leq y(t) \leq y_{\max} \nonumber  \\
   & \psi_{\min} \leq \psi(t) \leq \psi_{\max} \nonumber \\
   & \delta_{f, \min} \leq \delta_{f}(t) \leq \delta_{f, \max} \nonumber \\
   &U_{\min} \leq U(t) \leq U_{\max}  \nonumber \\
   &\gamma_{f, \min} \leq \gamma_{f} \left( t \right) \leq \gamma_{f, \max} \nonumber \\
   &J_{x, \min} \leq J_{x} \left( t \right) \leq J_{x, \max}\nonumber
\end{align*}

Finally, nonlinear functions of the vehicle's speed bound the vehicle's acceleration as
\begin{align*}
   & a_{x, \min}[U(t)]  \leq a_{x}(t) \leq a_{x, \max}[U(t)]  \nonumber
\end{align*}
Maximum deceleration/acceleration data collected from a $14$DoF HMMWV model are used to establish these nonlinear functions for the maximum deceleration/acceleration \cite{liu2017combined}.

No explicit lateral speed or yaw rate constraints exist.

\paragraph{Obstacle Avoidance}
Two possible approaches for incorporating the static and moving obstacle avoidance specifications into the OCP include soft constraints (or artificial potential-fields) and time-varying hard constraints \cite{zhang2017optimization}. There are two limitations to the soft constraints approach: (1) a trajectory may be generated that is deemed feasible according to the formulation, but actually goes through an obstacle, and (2), the NLP solve-times are known to be large, when compared with the time-varying hard constraints approach \cite{febbo2017moving}. Therefore, in this formulation, time-varying hard constraints for the avoidance of static and moving obstacles avoidance are incorporated into Eqn. \ref{eqn:path}.

Time-varying hard constraints enforce the vehicle's trajectory to avoid intersecting with the obstacles' trajectories, while accounting for the obstacles' shapes and sizes. Because this OCP will be transcribed into an NLP, the obstacles' shapes should be represented with twice continuously differentiable functions, e.g., a circle or an ellipse. As such, similar to planners tailored for spacecraft \cite{jewison2015model} and UGV \cite{febbo2017moving} applications, this work establishes a moving obstacle avoidance specification using time-varying, elliptical hard constraints. Eqn. \ref{eqn:obs} defines these constraints as
\begin{align}
  & ({\frac{{x(t) - (\mathbf{x0_{obs}}[i] + \mathbf{v_x}t)   } }{{\mathbf{a_{obs}}[i] + sm(t)}}})^2 + \nonumber \\
  & ({\frac{{y(t) - (\mathbf{y0_{obs}}[i] + \mathbf{v_y}t) }}{{\mathbf{b_{obs}}[i] + sm(t)}}})^2 > 1, \; \text{for}\; i \in 1:Q \label{eqn:obs}
\end{align}
where $sm(t)=sm_1+\frac{sm_2-sm_1}{t_f}t$ describes the time-varying safety margin, which enforces the vehicle to operate further from the obstacles as $t$ increases, and $Q$ is the total number of obstacles. The notation $\mathbf{x0_{obs}}[i]$ refers to the $i$th element of the $\mathbf{x0_{obs}}$ vector.

\paragraph{LiDAR Region Constraints} \label{sec:lidar}
To ensure that the vehicle's trajectory does not go beyond the LiDAR region, an additional path constraint is incorporated into Eqn. \ref{eqn:path}. This constraint is defined in Eqn. \ref{eqn:lidarA} as
\begin{equation}
(x(t)-x(t_0+t_{ex}))^2+(y(t)-y(t_0+t_{ex}))^2 - (L_{range} + \kappa)^2 \leq 0 \label{eqn:lidarA}
\end{equation}
where $\kappa$ is the LiDAR relaxation range \cite{liu2017combined}.

\paragraph{Initial and Terminal State Constraints} \label{sec:soft}
In a low-tolerance hard constraints approach, if the plant is driven into an infeasible state space, a feasible control signal cannot be computed \cite{scokaert1999feasibility}. To mitigate infeasible problems created using low-tolerance hard constraints on the initial and terminal conditions, soft constraints are introduced into this formulation. Soft constraints are introduced using slack variables, where the size of the slack variable corresponds to the respective constraint violation \cite{kerrigan2000soft}. These slack constraints are shown in Eqn. \ref{eqn:sl1}-\ref{eqn:sl6} as
\begin{align}
& \mathbf{X0}_p -\xi(t_0+t_{ex}) \leq \mathbf{x_0}_s  \label{eqn:sl1} \\
& \mathbf{X0}_p +\xi(t_0+t_{ex}) \geq  \mathbf{x_0}_s \label{eqn:sl2} \\
& x_g -x(t_f) \leq  \mathbf{x_f}_s[1]  \label{eqn:sl3} \\
& x_g + x(t_f) \geq  \mathbf{x_f}_s [1] \label{eqn:sl4} \\
& y_g -y(t_f) \leq  \mathbf{x_f}_s[2] \label{eqn:sl5} \\
& y_g + y(t_f) \geq  \mathbf{x_f}_s [2] \label{eqn:sl6}
\end{align}
where $\mathbf{x_0}_s$  is the $n_{st}$ dimensional vector of slack variables for the initial conditions, and $\mathbf{x_f}_s$ is the two dimensional vector of slack variables for the terminal conditions.

Adding slack variables to the cost functional reduces the size of the slack constraint violations. The weight for these slack variables is chosen to be large enough to keep the slack constraint close to zero. The $\varrho$ term in Eqn. \ref{eqn:cost} is now defined in Eqn. \ref{eqn:slackCost} as
\begin{equation}
      \varrho=\mathbf{w}_{s0}\mathbf{x_0}_s  + \mathbf{w}_{sf}\mathbf{x_f}_s \label{eqn:slackCost}
\end{equation}
where $\mathbf{w}_{s0}$ is a $1\times n_{st}$ dimensional vector of individual weight terms on the slack variables for the initial state constraints, and $\mathbf{w}_{sf}$ is a $1\times 2$ dimensional vector of individual weight terms  on the slack variables for the final state constraints.


When using only soft constraints, "optimal" trajectories are found that have initial and terminal states which are too far from their desired values. Adding high-tolerance hard constraints on the initial and terminal state conditions mitigates this issue. Thus, high-tolerance hard constraints are added to Eqn. \ref{eqn:event}, where the entire initial state is constrained to match $\mathbf{X0}_p$ within a specified tolerance $\mathbf{X0}_{tol}$. Eqn. \ref{eqn:X0} establishes these constraints as
\begin{align}
      \mathbf{X0}_p-\mathbf{X0}_{tol} \leq \boldsymbol{\xi}(t_0+t_{ex}) & \leq \mathbf{X0}_p+\mathbf{X0}_{tol}\label{eqn:X0}
\end{align}

Additionally, the vehicle's final $x$ and $y$ positions are constrained to be within the goal tolerance $\sigma$ using Eqn. \ref{eqn:xf} - Eqn. \ref{eqn:yf} as
\begin{align}
   x_g-\sigma & \leq x(t_f) \leq x_g+\sigma \label{eqn:xf}\\
   y_g-\sigma & \leq y(t_f) \leq y_g+\sigma \label{eqn:yf}
\end{align}

If the distance from the vehicle to the goal is greater than the vehicle's planning range $L_{\mathrm{range}}$, then the soft (Eqn. \ref{eqn:sl1} - Eqn. \ref{eqn:sl6}) and hard constraints (Eqn. \ref{eqn:xf} - Eqn. \ref{eqn:yf}) on the final conditions are relaxed. Setting the elements in $\mathbf{w}_{sf}$ to zero relaxes the soft constraints, and setting $\sigma$ to $10^6\;\si{m}$ relaxes the hard constraints. The remaining parameter modifications and additional constraints needed to relax the assumption that the goal is within the vehicle's planning range \cite{febbo2017moving} are now presented.

\paragraph{LiDAR Range Constraints}
If the goal is not within $L_{\mathrm{range}}$ of the vehicle, then the vehicle is constrained to arrive at the edge of the LiDAR region within a distance of $\kappa$ at $t_f$. This is accomplished using Eqn. \ref{eqn:l1} and Eqn. \ref{eqn:l2} as
\begin{align}
(x(t_f)-x(t_0+t_{ex}))^2 + (y(t_f)-y(t_0+t_{ex}))^2  - p1 \leq 0 \label{eqn:l1}\\
- (x(t_f)-x(t_0+t_{ex}))^2 - (y(t_f)-y(t_0+t_{ex}))^2 + p2 \leq 0 \label{eqn:l2}
\end{align}
where $p1$ and $p2$ are set to $(L_{\mathrm{range}} + \kappa)^2$ and $(L_{\mathrm{range}} - \kappa)^2$, respectively.

To avoid creating an infeasible problem while continuing to drive the UGV towards the goal, the goal constraints described in the previous section (i.e., Eqn. \ref{eqn:sl1} - Eqn. \ref{eqn:yf}) are relaxed and a new soft constraint is used. This soft constraint minimizes the squared distance from the vehicle to the goal at $t_f$, normalized by squared distance from the vehicle to the goal at $t_0 +t_{ex}$ \cite{liu2017combined}. Setting the goal weight $w_g$ in Eqn. \ref{eqn:cost_a} to a non-zero value enforces this constraint.

In the case that the goal is within a distance of $L_{\mathrm{range}}$ to the vehicle, $w_g$ is set to zero and the vehicle is constrained to reach the goal using Eqn. \ref{eqn:sl1} - Eqn. \ref{eqn:yf}. This is done by setting $\sigma$ to a much smaller goal tolerance, which enforces the hard constraints on reaching the goal, and setting the elements in the weight vector $\mathbf{w}_{sf}$ to large positive weights establishes soft constraint on reaching the goal through slack variables. Then, to avoid creating an infeasible problem, the hard constraints for reaching the edge of the LiDAR region at $t_f$ are relaxed. To do this, $p1$ and $p2$ are set to $10^{-6}$ and $-10^{-6}$, respectively,

The above specifies the details of the NMPC-based trajectory planning formulation with specifications \reqAnsp-\reqFnsp. The evaluation of this formulation as a function of its specifications follows.
\section{Evaluation Description} \label{sec:methods}
The next section presents comparisons among four planners within three different test environments, and evaluates the proposed planner's ability to improve both safety and performance without increasing solve-times. This section describes these planners and their test environments. Afterwards, the computer hardware platform used to produce the results presented in this paper and the software configuration under which \NLOpt is evaluated are described.

\subsection{Planners}
\begin{figure}
\captionof{table}{Planners compared in the work} \label{fig:planners}
\begin{minipage}[c]{\linewidth}
 \begin{center}
  \includegraphics[width=0.95\textwidth]{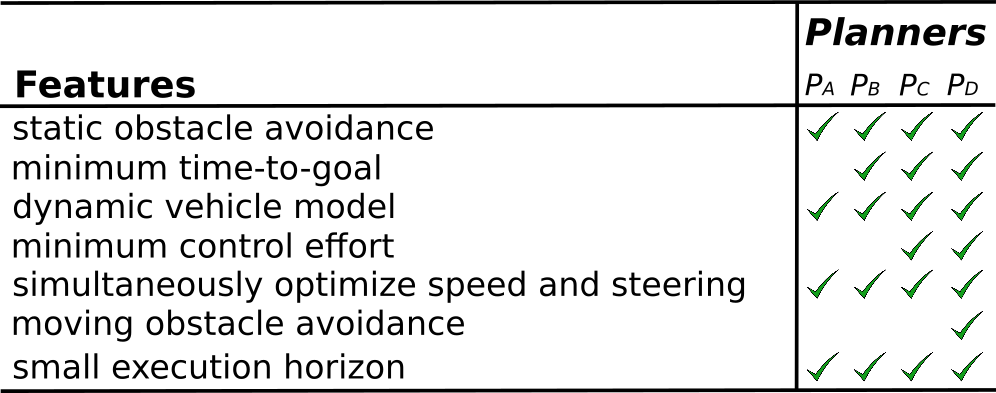}  
  \end{center}
\end{minipage}
\end{figure}

Comparisons are made among four planners (denoted as \plannerAnsp-\plannerD). The specifications of these planners are listed in Table \ref{fig:planners}, where \plannerA is used as the baseline planner. Note that \plannerA already includes the specifications of a dynamic vehicle model and simultaneous optimization of speed and steering, since previous work already illustrated the need to include them; see \cite{liu2017search,frasch2013auto,liu2016study} for the first and \cite{liu2017combined} for the latter.

The set of parameters in the left-hand column in Table \ref{tab:parama}, which is in Appendix \ref{appendixa}, define the planners. The right-hand column in Table \ref{tab:parama} defines the values of \plannerAnsp's parameters. All of the weight terms used in this work are obtained either from previous research \cite{liu2017combined} or manual tuning. In addition to this, setting the moving obstacle avoidance constraint in Eqn. \ref{eqn:obs} to $false$ means that Eqn. \ref{eqn:obs} is modified to Eqn. \ref{eqn:obs_b} as
\begin{equation}
  ({\frac{{x(t) - \mathbf{x0_{obs}\star}[i]}}{{\mathbf{a_{obs}}[i] + sm(t)}}})^2 + ({\frac{{y(t) - \mathbf{y0_{obs}\star}[i] }}{{\mathbf{b_{obs}}[i] + sm(t)}}})^2 > 1, \; \text{for}\; i \in 1:Q \label{eqn:obs_b}
\end{equation}
where $\mathbf{x0_{obs}\star}$ and $\mathbf{y0_{obs}\star}$ are arrays that describe the initial $x$ and $y$ positions of the obstacles, respectively. These arrays are updated to reflect the obstacles' current position after each execution horizon.

Next, the values of \plannerAnsp's parameters are modified to define \plannerBnsp-\plannerDnsp. The difference between \plannerA and \plannerB is that the weight on the final time $w_t$ is set to $100$ for \plannerBnsp.
To allow \plannerA to reach the goal or the edge of the LiDAR region at will, the final time is left as a design variable, but the weight on it (i.e., $w_t$) is set to zero. Next, the difference between \plannerB and \plannerC is that $w_{ce}$ is set to $1$ for the latter. Finally, the difference between \plannerC and \plannerD is that \plannerD has a moving obstacle avoidance specification, while \plannerC does not. \plannerD establishes this specification with Eqn. \ref{eqn:obs}. Specifically, \plannerC assumes that the obstacles will be static over each prediction horizon, while \plannerD incorporates the movement of the obstacles into the position constraints over the prediction horizon.

Closed-loop comparisons are made among four different vehicles (denoted as $V_A-V_D$). \plannerAnsp, \plannerBnsp, \plannerCnsp, and \plannerD control $V_A$, $V_B$, $V_C$, and $V_D$, respectively. Unless otherwise noted, the execution horizon is set to a value of $0.5\;\si{s}$ for all comparisons.

\subsection{Environment Categories}
To evaluate the proposed planner, this work uses four distinct environmental categories: unknown vs. known; unstructured vs. structured; dynamic vs. static, and challenging vs. simple. A description of each follows.

\paragraph{Unknown vs. Known}
In an unknown environment \cite{liu2017combined}, sensors collect data from the environment for algorithms that estimate factors including the obstacles' sizes, positions, and velocities. In addition to assuming that the obstacle information is known, this paper assumes that the environment known.

\paragraph{Unstructured vs. Structured}
In an unstructured environment \cite{thrun2006stanley,dahlkamp2006self,krusi2017driving,liu2017combined}, there are no roads to follow or traffic rules to obey. However, in a structured environment \cite{katrakazas2015real,paden2016survey}, some combination of these factors needs to be considered. In this paper the distinction between these environment categories is that in an unstructured environment there are no lanes to follow, while in a structured environment there is a lane to follow.

\paragraph{Dynamic vs. Static}
In a dynamic environment \cite{chiangsafety}, at least one obstacle is moving. In a static environment, all of the obstacles are stationary. This paper uses both dynamic and static environments.

\paragraph{Challenging vs. Simple}
In planning problems, the number of obstacles directly affects the computational load \cite{chiangsafety}. As such, the environment becomes more challenging as the number of obstacles increases. This paper uses an environment with $38$ obstacles as a challenging example and an environment with $3$ obstacles as a simple one.

\subsection{Environments}
UGV safety, performance, and solve-times are evaluated in three different environments (denoted as \caseAnsp-\caseCnsp). Each of these environments consists of some combination of the above environment categories, which is now described in detail.

\paragraph{\caseAnsp: Simple, Static, Unstructured Environment}
Both the increase in performance and solve-times, consequent to including the minimum time-to-goal and minimum control effort specifications, can be evaluated in a simple, static, unstructured environment (denoted as \caseAnsp). \caseA has three static obstacles (denoted as $0_1$, $0_2$, and $0_3$) and Table \ref{tab:enva}, which is in Appendix \ref{appendixa}, lists \caseAnsp's parameters. The right trace of Fig. \ref{fig:mt} shows the obstacle field and goal location of \caseAnsp.

\paragraph{\caseBnsp: Simple, Dynamic, Unstructured Environment}
The increases in both safety and solve-times, consequent to including a moving obstacle avoidance specification, can be evaluated in a simple, dynamic, unstructured environment (denoted as \caseBnsp). \caseB has three dynamic obstacles (also denoted as $0_1$, $0_2$, and $0_3$). \caseB is the same environment as \caseAnsp, except the obstacles are given non-zero velocities to test the planner's ability to avoid collisions with moving obstacles. The respective velocities of $0_1$, $0_2$, and $0_3$ are as follows:
$$
\mathbf{v_x} = [-2,-1,-0.5]\; \frac{\si{m}}{\si{s}}\;\mathrm{and}\;
\mathbf{v_y} = [0,1,6]\; \frac{\si{m}}{\si{s}}
$$

Fig. \ref{fig:mo} shows the movement of these obstacles, which can be seen in the right trace by following the obstacles' position at the indicated times.

\paragraph{\caseCnsp: Challenging, Dynamic, Structured Environment}
Increases in safety and solve-times from including a moving obstacle avoidance specification can be further evaluated within a challenging, dynamic, structured environment (denoted as \caseCnsp). \caseC is a double lane change scenario, which was originally developed to test a HMMWV within a static environment \cite{macadam1988development}. Fig. \ref{fig:moa} shows \caseCnsp, which has two large obstacles that need to be avoided, labeled $0_1$ and $0_2$. In this test, the vehicle is started at the bottom of Fig. \ref{fig:moa} traveling in the left lane at a speed of $17\;\frac{\si{m}}{\si{s}}$. From this point, it is restricted to perform a double lane change maneuver. To constrain the vehicle to perform this maneuver, first, minimum and maximum constraints on the vehicle's $x$ position are imposed. This restricted region is colored in light blue, where the lower and upper limits on the vehicle's $x$ position are $x_{min} = 0\;\si{m}$ and $x_{max}=24\;\si{m}$, respectively. Next, to ensure that the vehicle stays in the left lane during the first part of the maneuver, at the start of the track, until after $y=175\;\si{m}$, a series of $36$ cones are placed at the edge of the lane boundary. If these cones are not present, or there are not enough cones, then the vehicle will change lanes earlier in order to minimize the sixth term in Eqn. \ref{eqn:cost_a}. The first large obstacle, $0_1$, is static and is located in the left lane. The second large obstacle, $0_2$, starts at the back of the track in the right lane near the goal and moves towards the front of the track. Table \ref{tab:envc} lists the parameters for \caseCnsp.

To improve safety and performance within \caseCnsp, several planner parameters, which are listed in Table \ref{tab:parama}, are modified. Specifically, for all of the \caseC simulations shown in this paper, the $L_\text{range}$, $N$, and $\kappa$ planner parameters in Table \ref{tab:parama} are modified. The $L_\text{range}$ is modified because using an $L_\text{range}$ of $50\;\si{m}$, the vehicle crashes into $0_2$ over a large range of $0_2$ obstacle speeds. To avoid limiting the UGV based on its sensing range, and not its dynamic limits, $L_\text{range}$ is increased to $90\;\si{m}$ for \caseCnsp. To accommodate for this extended planning range, the number of points in the discretization $N$ is increased from $10$ to $15$ and the LiDAR relaxation range $\kappa$ is increased from $5\;\si{m}$ to $10 \;\si{m}$.

\subsection{Hardware Platform and Software Stack}
The results in this paper are produced using a single machine running Ubuntu $16.04$ with an Intel Core $\mathrm{i}7-4910\mathrm{MQ}$ CPU $@ 2.90\mathrm{GHz} \times 8$, and $31.3\mathrm{GB}$ of $\mathrm{RAM}$. This work evaluates \NLOptnsp's ability to solve the complex OCP formulation presented in this work in real-time. As mentioned, \NLOpt is our open-source, direct-collocation based OCP solver. In this work \NLOpt $0.1.6$ \cite{febbo_2017} is used with the \KNITRO $10.3$ NLP solver, where the default \KNITRO settings are used, except the maximum CPU-time (i.e., solve-time), which is set to $300\;\si{s}$. Additionally, the trapezoidal method \cite{Betts2010,kelly2017introduction} is used to approximate both the cost functional (Eqn. \ref{eqn:cost}) and the dynamics (Eqn. \ref{eqn:dynamics}). In order to more closely simulate practice, where time can typically be allocated to initialize trajectory planners, the optimizations are warm-started.
\section{Results} \label{sec:resultsB}
\subsection{Performance and Solve-Times within \caseA} \label{sec:ra}
Planning with a minimum time-to-goal specification can reduce the time-to-goal without increasing the solve-times in a simple, static, unstructured environment. In particular, using either \plannerB or \plannerC in lieu of \plannerA within \caseAnsp, reduces the time-to-goal from $9.0\;\si{s}$ to $7.0\;\si{s}$ (see Fig. \ref{fig:mt}). This is because, until about $t=5.5\;\si{s}$, both $V_B$ and $V_C$ accelerate while $V_A$ decelerates; this results in higher speeds for both $V_B$ and $V_C$. Next, while both \plannerB and \plannerC run in real-time in \caseAnsp, \plannerA does not. This can be seen in the top left trace of Fig. \ref{fig:mt}, where the solve-times for both \plannerB and \plannerC are all less than $t_{ex}$, but several of the solve-times obtained using \plannerA go above $t_{ex}$. Again, this creates a safety issue because, in practice, if the solve-time is greater than $t_{ex}$, then the vehicle will not have a trajectory to follow.

\pgfplotstableread{res/s3_A2/results/plant.csv}{\A}
\pgfplotstableread{res/s3_B2/results/plant.csv}{\B}
\pgfplotstableread{res/s3_C2/results/plant.csv}{\C}

\pgfplotstableread{res/s3_A2/results/opt.csv}{\optA}
\pgfplotstableread{res/s3_B2/results/opt.csv}{\optB}
\pgfplotstableread{res/s3_C2/results/opt.csv}{\optC}

\begin{filecontents*}{res/s3_B2/results/case.csv}
wgoal,wpsi,wtime,whaf,wce,csa,wsr,wjx,path,driver,ixt,iyt,ivt,irt,ipsit,isat,iut,iaxt,fxt,fyt,fvt,frt,fpsit,fsat,fut,faxt,outlev,feastolAbs,maxit,maxtimeCpu,ftol,feastol,infeastol,maxfevals,maxtimeReal,opttol,opttolAbs,xtol,acceptableObjChangeTol,warmStartInitPoint,dualInfTol,acceptableTol,acceptableConstrViolTol,acceptableDualInfTol,acceptableComplInfTol,divergingIteratesTol,radius,sX,sy,Xi,Yi,xRef,yRef,psiRef,goalTol,xi,yi,vi,ri,psii,sai,ui,axi,model,Xmin,Xmax,Ymin,Ymax,tp,tex,sms,smh,Lr,Lrd,NI,colPts,Nck,solver,MPCmaxIter,predictX0,fixedTp,integrationScheme,tfMax
10.0,0.01,100.0,1.0,0.0,0.1,1.0,0.01,1.0,1.0,0.5,0.5,0.5,0.005,0.5,0.025,0.5,0.5,5.0,5.0,NaN,NaN,NaN,NaN,NaN,NaN,0,0.07,500,30.0,1.0e-15,1.0e20,0.01,-1,30.0,1.0e20,0.5,1.0e-12,1.0e20,yes,5.0,0.01,0.01,1.0e10,0.01,1.0e20, [5.0\ 10.0\ 2.0] , [0.0\ 0.0\ 0.0] , [0.0\ 0.0\ 0.0] , [205.0\ 180.0\ 200.0] , [50.0\ 75.0\ 65.0] ,200.0,125.0,1.5701,15.0,200.0,0.0,0.0,0.0,1.57079632679,0.0,17.0,0.0,ThreeDOFv2,NaN,NaN,NaN,NaN,NaN,0.5,2.0,4.0,50.0,5.0,NaN,10,NaN,KNITRO,300,true,true,trapezoidal,10
\end{filecontents*}
\DTLloaddb[noheader=false]{case}{res/s3_B2/results/case.csv}
\begin{filecontents*}{res/s3_A2/results/veh.csv}
idx,t,xv,yv,psi
1,0.0,200.0,0.0,1.57079632679
2,3.0,196.5333983649164,49.55325676905725,1.7428678792753531
3,6.0,199.0518595697143,90.17853660632964,1.359505520988025
4,9.0,199.9147885878416,116.23949272486628,1.6016346138224757
\end{filecontents*}
\DTLloaddb[noheader=false]{vehA}{res/s3_A2/results/veh.csv}
\begin{filecontents*}{res/s3_B2/results/veh.csv}
idx,t,xv,yv,psi
1,0.0,200.0,0.0,1.57079632679
2,2.2,198.53138592344644,37.22625672841762,1.7171925623676625
3,4.3,193.9023058133953,78.42998213825712,1.4983160066284307
4,6.5,199.95791700360834,120.59558146566579,1.4122472598470928
\end{filecontents*}
\DTLloaddb[noheader=false]{vehB}{res/s3_B2/results/veh.csv}
\begin{filecontents*}{res/s3_C2/results/veh.csv}
idx,t,xv,yv,psi
1,0.0,200.0,0.0,1.57079632679
2,2.3,198.26261811403683,34.26640093007628,1.697808209014234
3,4.7,195.04497029967962,77.67546151534401,1.4692509884436418
4,7.0,200.99263244919777,124.10771517918585,1.5173221807351733
\end{filecontents*}
\DTLloaddb[noheader=false]{vehC}{res/s3_C2/results/veh.csv}

\begin{filecontents*}{res/s3_B2/results/obs_1.csv}
idx,t,xo,yo,ro,xv,yv,clr
1,0.0,205.0,50.0,5.0,200.0,0.0,0.0
2,2.2,205.0,50.0,5.0,198.53138592344644,37.22625672841762,25.0
3,4.3,205.0,50.0,5.0,193.9023058133953,78.42998213825712,50.0
4,6.5,205.0,50.0,5.0,199.95791700360834,120.59558146566579,75.0
\end{filecontents*}
\DTLloaddb[noheader=false]{obs_1}{res/s3_B2/results/obs_1.csv}
\begin{filecontents*}{res/s3_B2/results/obs_2.csv}
idx,t,xo,yo,ro,xv,yv,clr
1,0.0,180.0,75.0,10.0,200.0,0.0,0.0
2,2.2,180.0,75.0,10.0,198.53138592344644,37.22625672841762,25.0
3,4.3,180.0,75.0,10.0,193.9023058133953,78.42998213825712,50.0
4,6.5,180.0,75.0,10.0,199.95791700360834,120.59558146566579,75.0
\end{filecontents*}
\DTLloaddb[noheader=false]{obs_2}{res/s3_B2/results/obs_2.csv}
\begin{filecontents*}{res/s3_B2/results/obs_3.csv}
idx,t,xo,yo,ro,xv,yv,clr
1,0.0,200.0,65.0,2.0,200.0,0.0,0.0
2,2.2,200.0,65.0,2.0,198.53138592344644,37.22625672841762,25.0
3,4.3,200.0,65.0,2.0,193.9023058133953,78.42998213825712,50.0
4,6.5,200.0,65.0,2.0,199.95791700360834,120.59558146566579,75.0
\end{filecontents*}
\DTLloaddb[noheader=false]{obs_3}{res/s3_B2/results/obs_3.csv}

\if\slides1
\newcommand{\heightA}{3cm}
\newcommand{\widthA}{7cm}

\newcommand{\heightB}{5.5cm}
\newcommand{\widthB}{6cm}
\else

 \if\thesis0
  \newcommand{\heightA}{3.6cm}
  \newcommand{\widthA}{7cm}

  \newcommand{\heightB}{11cm}
  \newcommand{\widthB}{7cm}
 \else
  \newcommand{\heightA}{3.6cm}
  \newcommand{\widthA}{6cm}

  \newcommand{\heightB}{11cm}
  \newcommand{\widthB}{4cm}
 \fi
\fi

\begin{figure*}

\begin{minipage}[b]{0.45\linewidth}
\begin{center}

\begin{tikzpicture}

\begin{groupplot}[group style={group name = myplot, group size = 1 by 4, horizontal sep = 0.4cm, vertical sep = 1cm}, height = \heightA, width = \widthA]

\nextgroupplot[
             xmin = 1,
   		     ylabel={solve-time $(\si{s})$},
             ylabel style={align=center, text width=2cm},
             xlabel={Iterations},
    	grid=major,
             cycle list name=my markers
             ]
\addplot +[mark=none] table [x=evalNum, y=tSolve] {\optA}; \label{plots:Amt}
\addplot +[mark=none] table [x=evalNum, y=tSolve] {\optB}; \label{plots:Bmt}
\addplot +[mark=none] table [x=evalNum, y=tSolve] {\optC}; \label{plots:Cmt}

\draw[dashed, very thick, blue] (0,0.5) -- (17,0.5);

\if\thesis0
  \node at (9,.35) [fill=white] {Real-time threshold $=0.5\;\si{s}$};
\fi

\nextgroupplot[
             xmin = 0,
             ylabel={$\delta_f(t) \;(\si{radian})$},
             ylabel style={align=center, text width=2cm},
    	grid=major,
    	xticklabels=\empty,
             cycle list name=my markers
             ]
\addplot +[mark=none] table [x=t, y=sa] {\A};
\addplot +[mark=none] table [x=t, y=sa] {\B};
\addplot +[mark=none] table [x=t, y=sa] {\C};

\nextgroupplot[
             xmin = 0,
             ylabel={ $u_x \;(\si{\frac{m}{s}})$},
             ylabel style={align=center, text width=2cm},
    	grid=major,
    	xticklabels=\empty,
             cycle list name=my markers
             ]
\addplot +[mark=none] table [x=t, y=ux] {\A};
\addplot +[mark=none] table [x=t, y=ux] {\B};
\addplot +[mark=none] table [x=t, y=ux] {\C};

\nextgroupplot[
             xmin = 0,
             ylabel={$a_x \;(\si{\frac{m}{s^2}})$},
   	ylabel style={align=center, text width=2cm},
   	xlabel={Time  $(\si{s})$},
    	grid=major,
             cycle list name=my markers
             ]
\addplot +[mark=none] table [x=t, y=ax] {\A};
\addplot +[mark=none] table [x=t, y=ax] {\B};
\addplot +[mark=none] table [x=t, y=ax] {\C};

\end{groupplot}  

\end{tikzpicture}
\end{center}
\end{minipage}
\begin{minipage}[b]{0.45\linewidth}
\begin{center}

\begin{tikzpicture}
\pgfplotsset{
    height = \heightB,
    ymin=0, ymax=138,
    xmin=150, xmax=245,
    scale only axis=true,
    axis equal image, 
    grid=major,
    ylabel={$y \;(\si{m})$},
    xlabel={$x \;(\si{m})$},
    cycle list name=my markers,
}
\coordinate (top) at (rel axis cs:0,1);
\begin{axis}[]

\DTLforeach*{case}{\xg=xRef, \yg=yRef, \goalTol=goalTol}{
\draw[white!50!green, fill] (\xg,\yg) circle (6);
\draw[->,black,very thick] (\xg+9,\yg+1) to (\xg+4,\yg);
\node [right] at (\xg+9,\yg+2) {Goal};

\addplot +[mark=none] table [x=x, y=y] {\A};
\addplot +[mark=none] table [x=x, y=y] {\B};
\addplot +[mark=none] table [x=x, y=y] {\C};
                                                }

\pgfplotsextra{
\DTLforeach*{obs_1}{\x=xo, \y=yo, \r=ro, \t=t}{
\draw[thick, fill=red] (\x,\y) circle (\r);
\node [below] at (\x,\y+1.5) {\textcolor{white}{$O_1$}};
}

\DTLforeach*{obs_2}{\x=xo, \y=yo, \r=ro, \t=t}{
\draw[thick, fill=red] (\x,\y) circle (\r);
\node [] at (\x,\y) {\textcolor{white}{$O_2$}};
}

\DTLforeach*{obs_3}{\x=xo, \y=yo, \r=ro, \t=t}{
\draw[thick, fill=red] (\x,\y) circle (\r);
\node [right] at (\x+2,\y) {$O_3$};
}

\DTLforeach*{vehA}{\xv=xv, \yv=yv, \psiV=psi, \t=t}{
\draw[blue, fill,rotate around={\psiV*180/pi:(\xv,\yv)}] (\xv-1.5,\yv-1) rectangle (\xv+1.5, \yv+1);
\node [above right] at (\xv+15,\yv-1) {$V_A \mathrm{,}\; \t$\;\si{s}};
                                                 }

\DTLforeach*{vehC}{\xv=xv, \yv=yv, \psiV=psi, \t=t}{
\draw[black, fill,rotate around={\psiV*180/pi:(\xv,\yv)}] (\xv-1.5,\yv-1) rectangle (\xv+1.5, \yv+1);
\node [above left] at (\xv-8,\yv+6) {$V_B \; \mathrm{and}\; V_C \mathrm{,}\;\t$\;\si{s}};
                                                 }

\DTLforeach*{vehC}{\xv=xv, \yv=yv, \psiV=psi, \t=t}{
\draw[orange, fill,rotate around={\psiV*180/pi:(\xv,\yv)}] (\xv-1.5,\yv-1) rectangle (\xv+1.5, \yv+1);
                                                 }

\DTLforeach*{vehA}{\xv=xv, \yv=yv, \psiV=psi, \t=t}{
\draw[->,black,very thick] (\xv+15,\yv+3) to (\xv,\yv);
                                                 }
\DTLforeach*{vehC}{\xv=xv, \yv=yv, \psiV=psi, \t=t}{
\draw[->,black,very thick] (\xv-8,\yv+8) to (\xv,\yv);
                                                 }

               } 

\end{axis}

\coordinate (bot) at (rel axis cs:1,0);

\path (top|-current bounding box.north)--
      coordinate(legendpos)
      (bot|-current bounding box.north);
\matrix[
	fill=white,
    matrix of nodes,
    anchor=south,
    draw,
    inner sep=0.2em,
    draw
  ]at([yshift=-10cm,xshift=-14ex]legendpos)
  {
    \ref{plots:Amt} & $V_A$ &[5pt]\\
    \ref{plots:Bmt} & $V_B$ & [5pt] \\
    \ref{plots:Cmt} & $V_C$ & [5pt] \\
   };
\end{tikzpicture}
\end{center}
\end{minipage}
\if\slides0
	\caption{Closed-loop comparison of \plannerAnsp, \plannerBnsp, and \plannerC in \caseAnsp.} \label{fig:mt}
\fi
\end{figure*}

In \caseAnsp, compared to \plannerBnsp, \plannerC reduces the control effort without increasing either the time-to-goal or the solve-times. More specifically, even though $V_B$ and $V_C$ arrive at the goal in $7.0\;\si{s}$, $V_C$ uses less control effort for all of the three control effort terms. The third term in Eqn. \ref{eqn:cost_a} calculates the control effort terms for the steering angle, steering rate,  and longitudinal jerk. The overall values of each of these control effort terms, along with their percentage decrease, are in Table \ref{tab:cer}, which is in Appendix \ref{appendixa}. Next, in the top left trace in Fig. \ref{fig:mt}, it can be seen that the solve-times for \plannerB and \plannerC are below the real-time threshold of $0.5\;\si{s}$.

\subsection{Safety and Solve-Times within \caseB} \label{sec:rb}
Planning with a moving obstacle avoidance specification can increase safety without increasing the solve-times in a simple, dynamic, unstructured environment. This is shown in the comparison between \plannerC and \plannerD within \caseB (see Fig. \ref{fig:mo}). At the start of this test, both vehicles accelerate and then turn in opposite directions: $V_C$ to the left and $V_D$ to the right. \plannerC tries to avoid $0_1$ to the left, which results in a crash at $t=3.5\;\si{s}$. On the other hand, by taking the obstacles' motion over the prediction horizon into account, \plannerD turns $V_D$ to the right. This allows $V_D$ to arrive safely at the goal at $t=6.5\;\si{s}$. Lastly, as seen in the top left trace in Fig. \ref{fig:mo}, the solve-times for both \plannerB and \plannerC are below the real-time threshold of $0.5\;\si{s}$.

\pgfplotstableread{res/s4_C2/results/plant.csv}{\static}
\pgfplotstableread{res/s4_D2/results/plant.csv}{\move}

\pgfplotstableread{res/s4_C2/results/opt.csv}{\optF}
\pgfplotstableread{res/s4_D2/results/opt.csv}{\optT}

\DTLdeletedb{case}
\begin{filecontents*}{res/s4_C2/results/case.csv}
wgoal,wpsi,wtime,whaf,wce,csa,wsr,wjx,path,driver,ixt,iyt,ivt,irt,ipsit,isat,iut,iaxt,fxt,fyt,fvt,frt,fpsit,fsat,fut,faxt,outlev,feastolAbs,maxit,maxtimeCpu,ftol,feastol,infeastol,maxfevals,maxtimeReal,opttol,opttolAbs,xtol,acceptableObjChangeTol,warmStartInitPoint,dualInfTol,acceptableTol,acceptableConstrViolTol,acceptableDualInfTol,acceptableComplInfTol,divergingIteratesTol,radius,sX,sy,Xi,Yi,xRef,yRef,psiRef,goalTol,xi,yi,vi,ri,psii,sai,ui,axi,model,Xmin,Xmax,Ymin,Ymax,tp,tex,sms,smh,Lr,Lrd,NI,colPts,Nck,solver,MPCmaxIter,predictX0,fixedTp,integrationScheme,tfMax
10.0,0.01,100.0,1.0,1.0,0.1,1.0,0.01,1.0,1.0,0.5,0.5,0.5,0.005,0.5,0.025,0.5,0.5,5.0,5.0,NaN,NaN,NaN,NaN,NaN,NaN,0,0.07,500,30.0,1.0e-15,1.0e20,0.01,-1,30.0,1.0e20,0.5,1.0e-12,1.0e20,yes,5.0,0.01,0.01,1.0e10,0.01,1.0e20, [5.0\ 4.0\ 2.0] , Real[-2\ -1\ -0.5] , Real[0\ 1\ 6.0] , [205\ 180\ 200] , [57\ 75\ 63] ,200.0,125.0,1.5701,15.0,200.0,0.0,0.0,0.0,1.57079632679,0.0,17.0,0.0,ThreeDOFv2,NaN,NaN,NaN,NaN,NaN,0.5,2.0,4.0,50.0,5.0,NaN,10,NaN,KNITRO,300,true,true,trapezoidal,10
\end{filecontents*}
\DTLloaddb[noheader=false]{case}{res/s4_C2/results/case.csv}
\begin{filecontents*}{res/s4_C2/results/veh.csv}
idx,t,xv,yv,psi
1,0.0,200.0,0.0,1.57079632679
2,1.2,199.8931979758914,13.04756851014785,1.5971297553380053
3,2.3,198.06093710913635,34.23622385885485,1.7470499994234827
4,3.5,193.48070845222796,55.15251797392122,1.7896678652803923
\end{filecontents*}
\DTLloaddb[noheader=false]{vehF}{res/s4_C2/results/veh.csv}
\begin{filecontents*}{res/s4_D2/results/veh.csv}
idx,t,xv,yv,psi
1,0.0,200.0,0.0,1.57079632679
2,2.2,202.3728182287645,37.3743715880488,1.3384617378458392
3,4.3,210.64696728435484,78.23489064940891,1.548086830943054
4,6.5,203.99660006891398,120.12010627284965,1.9761473427449763
\end{filecontents*}
\DTLloaddb[noheader=false]{vehT}{res/s4_D2/results/veh.csv}

\DTLdeletedb{obs_1}
\DTLdeletedb{obs_2}
\DTLdeletedb{obs_3}
\begin{filecontents*}{res/s4_D2/results/obs_1.csv}
idx,t,xo,yo,ro,xv,yv,clr
1,0,205,57,5,200,0,100
2,2.2,200.6666666667,57,5,202.3728182288,37.374371588,66
3,4.3,196.3333333333,57,5,210.6469672844,78.2348906494,33
4,6.5,192,57,5,203.9966000689,120.1201062729,0
\end{filecontents*}
\DTLloaddb[noheader=false]{obs_1}{res/s4_D2/results/obs_1.csv}
\begin{filecontents*}{res/s4_D2/results/obs_2.csv}
idx,t,xo,yo,ro,xv,yv,clr
1,0,180,75,4,200,0,100
2,2.2,177.8333333333,77.1666666667,4,202.3728182288,37.374371588,66
3,4.3,175.6666666667,79.3333333333,4,210.6469672844,78.2348906494,33
4,6.5,173.5,81.5,4,203.9966000689,120.1201062729,0
\end{filecontents*}
\DTLloaddb[noheader=false]{obs_2}{res/s4_D2/results/obs_2.csv}
\begin{filecontents*}{res/s4_D2/results/obs_3.csv}
idx,t,xo,yo,ro,xv,yv,clr
1,0,200,63,2,200,0,100
2,2.2,198.9166666667,76,2,202.3728182288,37.374371588,66
3,4.3,197.8333333333,89,2,210.6469672844,78.2348906494,33
4,6.5,196.75,102,2,203.9966000689,120.1201062729,0
\end{filecontents*}
\DTLloaddb[noheader=false]{obs_3}{res/s4_D2/results/obs_3.csv}

\if\slides1
\newcommand{\heightAa}{3cm}
\newcommand{\widthAa}{7cm}

\newcommand{\heightBa}{5.5cm}
\newcommand{\widthBa}{6cm}
\else
             \if\thesis0
                          \newcommand{\heightAa}{3.6cm}
                          \newcommand{\widthAa}{7cm}

                          \newcommand{\heightBa}{11cm}
                          \newcommand{\widthBa}{7cm}
             \else
                          \newcommand{\heightAa}{3.6cm}
                          \newcommand{\widthAa}{6cm}

                          \newcommand{\heightBa}{11cm}
                          \newcommand{\widthBa}{4cm}
             \fi
\fi

\begin{figure*}

\begin{minipage}[b]{0.45\linewidth}
\begin{center}

\begin{tikzpicture}

\begin{groupplot}[group style={group name = myplot, group size = 1 by 4, horizontal sep = 0.4cm, vertical sep = 1cm}, height = \heightAa, width = \widthAa]

\nextgroupplot[
             xmin = 1,
		     ymax = 1,
                               ymax = 0.6,
   		     ylabel={solve-time $(\si{s})$},
             ylabel style={align=center, text width=2cm},
             xlabel={Iterations},
    	grid=major,
             cycle list name=my markers
             ]
\addplot +[mark=none] table [x=evalNum, y=tSolve] {\optT};
\addplot +[mark=none] table [x=evalNum, y=tSolve] {\optF};

\draw[dashed, very thick, blue] (0,0.5) -- (17,0.5);
\if\thesis0
\node at (7,0.35) [fill=white] {Real-time threshold $=0.5\;\si{s}$};
\fi
\nextgroupplot[
             xmin = 0,
             ylabel={$\delta_f(t)\;(\si{radian})$},
             ylabel style={align=center, text width=2cm},
    	grid=major,
    	xticklabels=\empty,
             cycle list name=my markers
             ]
\addplot +[mark=none] table [x=t, y=sa] {\move}; \label{plots:Dmmm}
\addplot +[mark=none] table [x=t, y=sa] {\static};  \label{plots:Cmmm}

\nextgroupplot[
             xmin = 0,
             ylabel={ $u_x\;(\si{\frac{m}{s}})$},
             ylabel style={align=center, text width=2cm},
    	grid=major,
    	xticklabels=\empty,
             cycle list name=my markers
             ]
\addplot +[mark=none] table [x=t, y=ux] {\move};
\addplot +[mark=none] table [x=t, y=ux] {\static};

\nextgroupplot[
             xmin = 0,
             ylabel={$a_x\;(\si{\frac{m}{s^2}})$},
   	ylabel style={align=center, text width=2cm},
   	xlabel={Time $(\si{s})$},
    	grid=major,
             cycle list name=my markers
             ]
\addplot +[mark=none] table [x=t, y=ax] {\move};
\addplot +[mark=none] table [x=t, y=ax] {\static};

\end{groupplot}  

\end{tikzpicture}
\end{center}
\end{minipage}
\begin{minipage}[b]{0.45\linewidth}
\begin{center}

\begin{tikzpicture}
\pgfplotsset{
    height = \heightBa,
    ymin=0, ymax=135,
    xmin=140, xmax=245,
    scale only axis=true,
    axis equal image, 
    grid=major,
    ylabel={$y\;(\si{m})$},
    xlabel={$x\;(\si{m})$},
    cycle list name=my markers,
}
\coordinate (top) at (rel axis cs:0,1);
\begin{axis}[]

\DTLforeach*{case}{\xg=xRef, \yg=yRef, \goalTol=goalTol}{
\draw[white!50!green, fill] (\xg,\yg) circle (5);
\draw[->,black,very thick] (\xg-9,\yg-2) to (\xg,\yg);
\node [left] at (\xg-9,\yg-2) {Goal};

\addplot +[mark=none] table [x=x, y=y] {\move};
\addplot +[mark=none] table [x=x, y=y] {\static};
                                                }

\pgfplotsextra{
\DTLforeach*{obs_1}{\x=xo, \y=yo, \r=ro, \t=t,\clr=clr}{
\draw[thick, fill=pink!\clr!red] (\x,\y) circle (\r);
}
\DTLforeach*{obs_1}{\x=xo, \y=yo, \r=ro, \t=t, \idx=idx}{
\draw[->,black,very thick] (\x+10+\idx*6,\y+10-\idx*5) to (\x+2,\y);
\node [right] at (\x+10+\idx*6,\y+10-\idx*5) {$O_1 \mathrm{,}\; \t$\;\si{s}};
}

\DTLforeach*{obs_2}{\x=xo, \y=yo, \r=ro, \t=t,\clr=clr}{
\draw[thick, fill=pink!\clr!red] (\x,\y) circle (\r);
}
\DTLforeach*{obs_2}{\x=xo, \y=yo, \r=ro, \t=t, \idx=idx}{
\draw[->,black,very thick]  (\x-10+\idx,\y-30+\idx*4) to (\x+1.2,\y-1.5);
\node [left] at (\x-10+\idx,\y-30+\idx*4) {$O_2 \mathrm{,}\; \t$\;\si{s}};
}

\DTLforeach*{obs_3}{\x=xo, \y=yo, \r=ro, \t=t,\clr=clr}{
\draw[thick, fill=pink!\clr!red] (\x,\y) circle (\r);
}
\DTLforeach*{obs_3}{\x=xo, \y=yo, \r=ro, \t=t}{
\node [above] at (\x,\y+0.75) {$O_3\mathrm{,}\; \t$\;s};
}
\DTLforeach*{vehF}{\xv=xv, \yv=yv, \psiV=psi, \t=t}{
\draw[blue, fill,rotate around={\psiV*180/pi:(\xv,\yv)}] (\xv-1.5,\yv-1) rectangle (\xv+1.5, \yv+1);
\node [above left] at (\xv-1,\yv) {$V_C \mathrm{,}\; \t$\;\si{s}};
                                                 }

\DTLforeach*{vehT}{\xv=xv, \yv=yv, \psiV=psi, \t=t}{
\draw[orange, fill,rotate around={\psiV*180/pi:(\xv,\yv)}] (\xv-1.5,\yv-1) rectangle (\xv+1.5, \yv+1);
\node [above left] at (\xv+17,\yv) {$V_D \mathrm{,}\;\t$\;s};
                                                 }

               } 

\end{axis}

\coordinate (bot) at (rel axis cs:1,0);

\path (top|-current bounding box.north)--
      coordinate(legendpos)
      (bot|-current bounding box.north);
\matrix[
	fill=white,
    matrix of nodes,
    anchor=south,
    draw,
    inner sep=0.2em,
    draw
  ]at([yshift=-10ex,xshift=-14ex]legendpos)
  {
    \ref{plots:Cmmm} & $V_C$ & [5pt] \\
    \ref{plots:Dmmm} & $V_D$ & [5pt] \\
   };
\end{tikzpicture}
\end{center}
\end{minipage}
\if\slides0
  \caption{Closed-loop comparison of \plannerC and \plannerD in \caseBnsp.  \label{fig:mo}}
\fi
\end{figure*}

\subsection{Safety and Solve-Times within \caseC} \label{sec:rc}
Similarly, planning with a moving obstacle avoidance specification can increase safety without significantly increasing the solve-times in a challenging, dynamic, structured environment. This is demonstrated by testing \plannerC and \plannerD within \caseC (see Fig. \ref{fig:moa} - Fig. \ref{fig:moc}). At the start of the test, the first lane-change maneuver is performed successfully for both  $V_C$ and $V_D$.

During this time, both vehicles accelerate aggressively to increase their speed from $17\;\frac{\si{m}}{\si{s}}$ at $t=0\;\si{s}$ to $26.5\;\frac{\si{m}}{\si{s}}$ at $t=19.5\;\si{s}$. At this time, $V_C$ crashes into $0_2$ (see Fig. \ref{fig:mob} for a zoomed in view of the crash) while $V_D$ avoids $0_2$ and eventually attains the goal. Additionally, \plannerD is able to avoid this collision with a solve-time that is only slightly higher than the one obtained with \plannerCnsp's just before is causes $V_C$ to crash. The next section discusses the larger solve-times encountered at $19.0\;\si{s}$. Finally, the solve-times for \plannerD are less than the real-time threshold of $0.5\;\si{s}$, despite the fact that this is a challenging environment (i.e., with $38$ obstacles instead of $3$).

\pgfplotstableread{res/s9_C2/results/plant.csv}{\static}
\pgfplotstableread{res/s9_D2/results/plant.csv}{\move}

\pgfplotstableread{res/s9_C2/results/opt.csv}{\optF}
\pgfplotstableread{res/s9_D2/results/opt.csv}{\optT}

\begin{filecontents*}{res/s9_C2/results/case.csv}
wgoal,wpsi,wtime,whaf,wce,csa,wsr,wjx,path,driver,ixt,iyt,ivt,irt,ipsit,isat,iut,iaxt,fxt,fyt,fvt,frt,fpsit,fsat,fut,faxt,outlev,feastolAbs,maxit,maxtimeCpu,ftol,feastol,infeastol,maxfevals,maxtimeReal,opttol,opttolAbs,xtol,acceptableObjChangeTol,warmStartInitPoint,dualInfTol,acceptableTol,acceptableConstrViolTol,acceptableDualInfTol,acceptableComplInfTol,divergingIteratesTol,radius,sX,sy,Xi,Yi,xRef,yRef,psiRef,goalTol,xi,yi,vi,ri,psii,sai,ui,axi,model,Xmin,Xmax,Ymin,Ymax,tp,tex,sms,smh,Lr,Lrd,NI,colPts,Nck,solver,MPCmaxIter,predictX0,fixedTp,integrationScheme,tfMax
10.0,0.01,100.0,1,1.0,0.1,1.0,0.01,1.0,1.0,0.5,0.5,0.5,0.005,0.5,0.025,0.5,0.5,5.0,5.0,NaN,NaN,NaN,NaN,NaN,NaN,0,0.07,500,30.0,1.0e-15,1.0e20,0.01,-1,30.0,1.0e20,0.5,1.0e-12,1.0e20,yes,5.0,0.01,0.01,1.0e10,0.01,1.0e20, [6.0\ 6.0\ 0.38735\ 0.38735\ 0.38735\ 0.38735\ 0.38735\ 0.38735\ 0.38735\ 0.38735\ 0.38735\ 0.38735\ 0.38735\ 0.38735\ 0.38735\ 0.38735\ 0.38735\ 0.38735\ 0.38735\ 0.38735\ 0.38735\ 0.38735\ 0.38735\ 0.38735\ 0.38735\ 0.38735\ 0.38735\ 0.38735\ 0.38735\ 0.38735\ 0.38735\ 0.38735\ 0.38735\ 0.38735\ 0.38735\ 0.38735\ 0.38735\ 0.38735] , [0.0\ 0.0\ 0.0\ 0.0\ 0.0\ 0.0\ 0.0\ 0.0\ 0.0\ 0.0\ 0.0\ 0.0\ 0.0\ 0.0\ 0.0\ 0.0\ 0.0\ 0.0\ 0.0\ 0.0\ 0.0\ 0.0\ 0.0\ 0.0\ 0.0\ 0.0\ 0.0\ 0.0\ 0.0\ 0.0\ 0.0\ 0.0\ 0.0\ 0.0\ 0.0\ 0.0\ 0.0\ 0.0] , [0.0\ -10.0\ 0.0\ 0.0\ 0.0\ 0.0\ 0.0\ 0.0\ 0.0\ 0.0\ 0.0\ 0.0\ 0.0\ 0.0\ 0.0\ 0.0\ 0.0\ 0.0\ 0.0\ 0.0\ 0.0\ 0.0\ 0.0\ 0.0\ 0.0\ 0.0\ 0.0\ 0.0\ 0.0\ 0.0\ 0.0\ 0.0\ 0.0\ 0.0\ 0.0\ 0.0\ 0.0\ 0.0] , [6.0\ 18.0\ 12.0\ 12.0\ 12.0\ 12.0\ 12.0\ 12.0\ 12.0\ 12.0\ 12.0\ 12.0\ 12.0\ 12.0\ 12.0\ 12.0\ 12.0\ 12.0\ 12.0\ 12.0\ 12.0\ 12.0\ 12.0\ 12.0\ 12.0\ 12.0\ 12.0\ 12.0\ 12.0\ 12.0\ 12.0\ 12.0\ 12.0\ 12.0\ 12.0\ 12.0\ 12.0\ 12.0] , [281.0\ 650.0\ 0.0\ 5.0\ 10.0\ 15.0\ 20.0\ 25.0\ 30.0\ 35.0\ 40.0\ 45.0\ 50.0\ 55.0\ 60.0\ 65.0\ 70.0\ 75.0\ 80.0\ 85.0\ 90.0\ 95.0\ 100.0\ 105.0\ 110.0\ 115.0\ 120.0\ 125.0\ 130.0\ 135.0\ 140.0\ 145.0\ 150.0\ 155.0\ 160.0\ 165.0\ 170.0\ 175.0] ,18.0,700.0,1.5701,25.0,6.0,0.0,0.0,0.0,1.57079632679,0.0,17.0,0.0,ThreeDOFv2,0.0,24.0,NaN,NaN,NaN,0.5,2.0,4.0,90.0,10.0,NaN,15,NaN,KNITRO,300,true,true,trapezoidal,10
\end{filecontents*}
\DTLloaddb[noheader=false]{case9}{res/s9_C2/results/case.csv}
\begin{filecontents*}{res/s9_C2/results/veh.csv}
idx,t,xv,yv,psi
1,0.0,6.0,0.0,1.57079632679
2,6.5,8.00364184891656,125.87107508109118,1.5623151958161403
3,13.0,18.073086250031288,279.8906420427333,1.5943165649373687
4,19.5,14.660984161080721,448.51891544755733,1.7000541544318875
\end{filecontents*}
\DTLloaddb[noheader=false]{vehF9}{res/s9_C2/results/veh.csv}
\begin{filecontents*}{res/s9_D2/results/veh.csv}
idx,t,xv,yv,psi
1,0.0,6.0,0.0,1.57079632679
2,9.5,11.178666434744933,194.6708036977185,1.3796724417884072
3,19.5,10.182713912027248,446.8108193616263,1.7913899426031334
4,28.5,17.761257602671837,687.8623353384033,1.5512279376011837
\end{filecontents*}
\DTLloaddb[noheader=false]{vehT9}{res/s9_D2/results/veh.csv}

\begin{filecontents*}{res/s9_D2/results/obs_1.csv}
idx,t,xo,yo,ro,xv,yv
1,0.0,6.0,281.0,6.0,6.0,0.0
2,9.5,6.0,281.0,6.0,11.178666434744933,194.6708036977185
3,19.5,6.0,281.0,6.0,10.182713912027248,446.8108193616263
4,28.5,6.0,281.0,6.0,17.761257602671837,687.8623353384033
\end{filecontents*}
\DTLloaddb[noheader=false]{obs1}{res/s9_D2/results/obs_1.csv}
\begin{filecontents*}{res/s9_D2/results/obs_2.csv}
idx,t,xo,yo,ro,xv,yv,clr
1,0,18,650,6,6,0,100
2,9.5,18,555,6,11.1786664347,194.6708036977,66
3,19.5,18,455,6,10.182713912,446.8108193616,33
4,28.5,18,365,6,17.7612576027,687.8623353384,0
\end{filecontents*}
\DTLloaddb[noheader=false]{obs2}{res/s9_D2/results/obs_2.csv}

\begin{filecontents*}{res/s9_D2/results/obs_3.csv}
idx,t,xo,yo,ro,xv,yv
1,0.0,12.0,0.0,0.38735,6.0,0.0
2,9.5,12.0,0.0,0.38735,11.178666434744933,194.6708036977185
3,19.5,12.0,0.0,0.38735,10.182713912027248,446.8108193616263
4,28.5,12.0,0.0,0.38735,17.761257602671837,687.8623353384033
\end{filecontents*}
\DTLloaddb[noheader=false]{obs3}{res/s9_D2/results/obs_3.csv}
\begin{filecontents*}{res/s9_D2/results/obs_4.csv}
idx,t,xo,yo,ro,xv,yv
1,0.0,12.0,5.0,0.38735,6.0,0.0
2,9.5,12.0,5.0,0.38735,11.178666434744933,194.6708036977185
3,19.5,12.0,5.0,0.38735,10.182713912027248,446.8108193616263
4,28.5,12.0,5.0,0.38735,17.761257602671837,687.8623353384033
\end{filecontents*}
\DTLloaddb[noheader=false]{obs4}{res/s9_D2/results/obs_4.csv}
\begin{filecontents*}{res/s9_D2/results/obs_5.csv}
idx,t,xo,yo,ro,xv,yv
1,0.0,12.0,10.0,0.38735,6.0,0.0
2,9.5,12.0,10.0,0.38735,11.178666434744933,194.6708036977185
3,19.5,12.0,10.0,0.38735,10.182713912027248,446.8108193616263
4,28.5,12.0,10.0,0.38735,17.761257602671837,687.8623353384033
\end{filecontents*}
\DTLloaddb[noheader=false]{obs5}{res/s9_D2/results/obs_5.csv}
\begin{filecontents*}{res/s9_D2/results/obs_6.csv}
idx,t,xo,yo,ro,xv,yv
1,0.0,12.0,15.0,0.38735,6.0,0.0
2,9.5,12.0,15.0,0.38735,11.178666434744933,194.6708036977185
3,19.5,12.0,15.0,0.38735,10.182713912027248,446.8108193616263
4,28.5,12.0,15.0,0.38735,17.761257602671837,687.8623353384033
\end{filecontents*}
\DTLloaddb[noheader=false]{obs6}{res/s9_D2/results/obs_6.csv}
\begin{filecontents*}{res/s9_D2/results/obs_7.csv}
idx,t,xo,yo,ro,xv,yv
1,0.0,12.0,20.0,0.38735,6.0,0.0
2,9.5,12.0,20.0,0.38735,11.178666434744933,194.6708036977185
3,19.5,12.0,20.0,0.38735,10.182713912027248,446.8108193616263
4,28.5,12.0,20.0,0.38735,17.761257602671837,687.8623353384033
\end{filecontents*}
\DTLloaddb[noheader=false]{obs7}{res/s9_D2/results/obs_7.csv}
\begin{filecontents*}{res/s9_D2/results/obs_8.csv}
idx,t,xo,yo,ro,xv,yv
1,0.0,12.0,25.0,0.38735,6.0,0.0
2,9.5,12.0,25.0,0.38735,11.178666434744933,194.6708036977185
3,19.5,12.0,25.0,0.38735,10.182713912027248,446.8108193616263
4,28.5,12.0,25.0,0.38735,17.761257602671837,687.8623353384033
\end{filecontents*}
\DTLloaddb[noheader=false]{obs8}{res/s9_D2/results/obs_8.csv}
\begin{filecontents*}{res/s9_D2/results/obs_9.csv}
idx,t,xo,yo,ro,xv,yv
1,0.0,12.0,30.0,0.38735,6.0,0.0
2,9.5,12.0,30.0,0.38735,11.178666434744933,194.6708036977185
3,19.5,12.0,30.0,0.38735,10.182713912027248,446.8108193616263
4,28.5,12.0,30.0,0.38735,17.761257602671837,687.8623353384033
\end{filecontents*}
\DTLloaddb[noheader=false]{obs9}{res/s9_D2/results/obs_9.csv}
\begin{filecontents*}{res/s9_D2/results/obs_10.csv}
idx,t,xo,yo,ro,xv,yv
1,0.0,12.0,35.0,0.38735,6.0,0.0
2,9.5,12.0,35.0,0.38735,11.178666434744933,194.6708036977185
3,19.5,12.0,35.0,0.38735,10.182713912027248,446.8108193616263
4,28.5,12.0,35.0,0.38735,17.761257602671837,687.8623353384033
\end{filecontents*}
\DTLloaddb[noheader=false]{obs10}{res/s9_D2/results/obs_10.csv}
\begin{filecontents*}{res/s9_D2/results/obs_11.csv}
idx,t,xo,yo,ro,xv,yv
1,0.0,12.0,40.0,0.38735,6.0,0.0
2,9.5,12.0,40.0,0.38735,11.178666434744933,194.6708036977185
3,19.5,12.0,40.0,0.38735,10.182713912027248,446.8108193616263
4,28.5,12.0,40.0,0.38735,17.761257602671837,687.8623353384033
\end{filecontents*}
\DTLloaddb[noheader=false]{obs11}{res/s9_D2/results/obs_11.csv}
\begin{filecontents*}{res/s9_D2/results/obs_12.csv}
idx,t,xo,yo,ro,xv,yv
1,0.0,12.0,45.0,0.38735,6.0,0.0
2,9.5,12.0,45.0,0.38735,11.178666434744933,194.6708036977185
3,19.5,12.0,45.0,0.38735,10.182713912027248,446.8108193616263
4,28.5,12.0,45.0,0.38735,17.761257602671837,687.8623353384033
\end{filecontents*}
\DTLloaddb[noheader=false]{obs12}{res/s9_D2/results/obs_12.csv}
\begin{filecontents*}{res/s9_D2/results/obs_13.csv}
idx,t,xo,yo,ro,xv,yv
1,0.0,12.0,50.0,0.38735,6.0,0.0
2,9.5,12.0,50.0,0.38735,11.178666434744933,194.6708036977185
3,19.5,12.0,50.0,0.38735,10.182713912027248,446.8108193616263
4,28.5,12.0,50.0,0.38735,17.761257602671837,687.8623353384033
\end{filecontents*}
\DTLloaddb[noheader=false]{obs13}{res/s9_D2/results/obs_13.csv}
\begin{filecontents*}{res/s9_D2/results/obs_14.csv}
idx,t,xo,yo,ro,xv,yv
1,0.0,12.0,55.0,0.38735,6.0,0.0
2,9.5,12.0,55.0,0.38735,11.178666434744933,194.6708036977185
3,19.5,12.0,55.0,0.38735,10.182713912027248,446.8108193616263
4,28.5,12.0,55.0,0.38735,17.761257602671837,687.8623353384033
\end{filecontents*}
\DTLloaddb[noheader=false]{obs14}{res/s9_D2/results/obs_14.csv}
\begin{filecontents*}{res/s9_D2/results/obs_15.csv}
idx,t,xo,yo,ro,xv,yv
1,0.0,12.0,60.0,0.38735,6.0,0.0
2,9.5,12.0,60.0,0.38735,11.178666434744933,194.6708036977185
3,19.5,12.0,60.0,0.38735,10.182713912027248,446.8108193616263
4,28.5,12.0,60.0,0.38735,17.761257602671837,687.8623353384033
\end{filecontents*}
\DTLloaddb[noheader=false]{obs15}{res/s9_D2/results/obs_15.csv}
\begin{filecontents*}{res/s9_D2/results/obs_16.csv}
idx,t,xo,yo,ro,xv,yv
1,0.0,12.0,65.0,0.38735,6.0,0.0
2,9.5,12.0,65.0,0.38735,11.178666434744933,194.6708036977185
3,19.5,12.0,65.0,0.38735,10.182713912027248,446.8108193616263
4,28.5,12.0,65.0,0.38735,17.761257602671837,687.8623353384033
\end{filecontents*}
\DTLloaddb[noheader=false]{obs16}{res/s9_D2/results/obs_16.csv}
\begin{filecontents*}{res/s9_D2/results/obs_17.csv}
idx,t,xo,yo,ro,xv,yv
1,0.0,12.0,70.0,0.38735,6.0,0.0
2,9.5,12.0,70.0,0.38735,11.178666434744933,194.6708036977185
3,19.5,12.0,70.0,0.38735,10.182713912027248,446.8108193616263
4,28.5,12.0,70.0,0.38735,17.761257602671837,687.8623353384033
\end{filecontents*}
\DTLloaddb[noheader=false]{obs17}{res/s9_D2/results/obs_17.csv}
\begin{filecontents*}{res/s9_D2/results/obs_18.csv}
idx,t,xo,yo,ro,xv,yv
1,0.0,12.0,75.0,0.38735,6.0,0.0
2,9.5,12.0,75.0,0.38735,11.178666434744933,194.6708036977185
3,19.5,12.0,75.0,0.38735,10.182713912027248,446.8108193616263
4,28.5,12.0,75.0,0.38735,17.761257602671837,687.8623353384033
\end{filecontents*}
\DTLloaddb[noheader=false]{obs18}{res/s9_D2/results/obs_18.csv}
\begin{filecontents*}{res/s9_D2/results/obs_19.csv}
idx,t,xo,yo,ro,xv,yv
1,0.0,12.0,80.0,0.38735,6.0,0.0
2,9.5,12.0,80.0,0.38735,11.178666434744933,194.6708036977185
3,19.5,12.0,80.0,0.38735,10.182713912027248,446.8108193616263
4,28.5,12.0,80.0,0.38735,17.761257602671837,687.8623353384033
\end{filecontents*}
\DTLloaddb[noheader=false]{obs19}{res/s9_D2/results/obs_19.csv}
\begin{filecontents*}{res/s9_D2/results/obs_20.csv}
idx,t,xo,yo,ro,xv,yv
1,0.0,12.0,85.0,0.38735,6.0,0.0
2,9.5,12.0,85.0,0.38735,11.178666434744933,194.6708036977185
3,19.5,12.0,85.0,0.38735,10.182713912027248,446.8108193616263
4,28.5,12.0,85.0,0.38735,17.761257602671837,687.8623353384033
\end{filecontents*}
\DTLloaddb[noheader=false]{obs20}{res/s9_D2/results/obs_20.csv}
\begin{filecontents*}{res/s9_D2/results/obs_21.csv}
idx,t,xo,yo,ro,xv,yv
1,0.0,12.0,90.0,0.38735,6.0,0.0
2,9.5,12.0,90.0,0.38735,11.178666434744933,194.6708036977185
3,19.5,12.0,90.0,0.38735,10.182713912027248,446.8108193616263
4,28.5,12.0,90.0,0.38735,17.761257602671837,687.8623353384033
\end{filecontents*}
\DTLloaddb[noheader=false]{obs21}{res/s9_D2/results/obs_21.csv}
\begin{filecontents*}{res/s9_D2/results/obs_22.csv}
idx,t,xo,yo,ro,xv,yv
1,0.0,12.0,95.0,0.38735,6.0,0.0
2,9.5,12.0,95.0,0.38735,11.178666434744933,194.6708036977185
3,19.5,12.0,95.0,0.38735,10.182713912027248,446.8108193616263
4,28.5,12.0,95.0,0.38735,17.761257602671837,687.8623353384033
\end{filecontents*}
\DTLloaddb[noheader=false]{obs22}{res/s9_D2/results/obs_22.csv}
\begin{filecontents*}{res/s9_D2/results/obs_23.csv}
idx,t,xo,yo,ro,xv,yv
1,0.0,12.0,100.0,0.38735,6.0,0.0
2,9.5,12.0,100.0,0.38735,11.178666434744933,194.6708036977185
3,19.5,12.0,100.0,0.38735,10.182713912027248,446.8108193616263
4,28.5,12.0,100.0,0.38735,17.761257602671837,687.8623353384033
\end{filecontents*}
\DTLloaddb[noheader=false]{obs23}{res/s9_D2/results/obs_23.csv}
\begin{filecontents*}{res/s9_D2/results/obs_24.csv}
idx,t,xo,yo,ro,xv,yv
1,0.0,12.0,105.0,0.38735,6.0,0.0
2,9.5,12.0,105.0,0.38735,11.178666434744933,194.6708036977185
3,19.5,12.0,105.0,0.38735,10.182713912027248,446.8108193616263
4,28.5,12.0,105.0,0.38735,17.761257602671837,687.8623353384033
\end{filecontents*}
\DTLloaddb[noheader=false]{obs24}{res/s9_D2/results/obs_24.csv}
\begin{filecontents*}{res/s9_D2/results/obs_25.csv}
idx,t,xo,yo,ro,xv,yv
1,0.0,12.0,110.0,0.38735,6.0,0.0
2,9.5,12.0,110.0,0.38735,11.178666434744933,194.6708036977185
3,19.5,12.0,110.0,0.38735,10.182713912027248,446.8108193616263
4,28.5,12.0,110.0,0.38735,17.761257602671837,687.8623353384033
\end{filecontents*}
\DTLloaddb[noheader=false]{obs25}{res/s9_D2/results/obs_25.csv}
\begin{filecontents*}{res/s9_D2/results/obs_26.csv}
idx,t,xo,yo,ro,xv,yv
1,0.0,12.0,115.0,0.38735,6.0,0.0
2,9.5,12.0,115.0,0.38735,11.178666434744933,194.6708036977185
3,19.5,12.0,115.0,0.38735,10.182713912027248,446.8108193616263
4,28.5,12.0,115.0,0.38735,17.761257602671837,687.8623353384033
\end{filecontents*}
\DTLloaddb[noheader=false]{obs26}{res/s9_D2/results/obs_26.csv}
\begin{filecontents*}{res/s9_D2/results/obs_27.csv}
idx,t,xo,yo,ro,xv,yv
1,0.0,12.0,120.0,0.38735,6.0,0.0
2,9.5,12.0,120.0,0.38735,11.178666434744933,194.6708036977185
3,19.5,12.0,120.0,0.38735,10.182713912027248,446.8108193616263
4,28.5,12.0,120.0,0.38735,17.761257602671837,687.8623353384033
\end{filecontents*}
\DTLloaddb[noheader=false]{obs27}{res/s9_D2/results/obs_27.csv}
\begin{filecontents*}{res/s9_D2/results/obs_28.csv}
idx,t,xo,yo,ro,xv,yv
1,0.0,12.0,125.0,0.38735,6.0,0.0
2,9.5,12.0,125.0,0.38735,11.178666434744933,194.6708036977185
3,19.5,12.0,125.0,0.38735,10.182713912027248,446.8108193616263
4,28.5,12.0,125.0,0.38735,17.761257602671837,687.8623353384033
\end{filecontents*}
\DTLloaddb[noheader=false]{obs28}{res/s9_D2/results/obs_28.csv}
\begin{filecontents*}{res/s9_D2/results/obs_29.csv}
idx,t,xo,yo,ro,xv,yv
1,0.0,12.0,130.0,0.38735,6.0,0.0
2,9.5,12.0,130.0,0.38735,11.178666434744933,194.6708036977185
3,19.5,12.0,130.0,0.38735,10.182713912027248,446.8108193616263
4,28.5,12.0,130.0,0.38735,17.761257602671837,687.8623353384033
\end{filecontents*}
\DTLloaddb[noheader=false]{obs29}{res/s9_D2/results/obs_29.csv}
\begin{filecontents*}{res/s9_D2/results/obs_30.csv}
idx,t,xo,yo,ro,xv,yv
1,0.0,12.0,135.0,0.38735,6.0,0.0
2,9.5,12.0,135.0,0.38735,11.178666434744933,194.6708036977185
3,19.5,12.0,135.0,0.38735,10.182713912027248,446.8108193616263
4,28.5,12.0,135.0,0.38735,17.761257602671837,687.8623353384033
\end{filecontents*}
\DTLloaddb[noheader=false]{obs30}{res/s9_D2/results/obs_30.csv}
\begin{filecontents*}{res/s9_D2/results/obs_31.csv}
idx,t,xo,yo,ro,xv,yv
1,0.0,12.0,140.0,0.38735,6.0,0.0
2,9.5,12.0,140.0,0.38735,11.178666434744933,194.6708036977185
3,19.5,12.0,140.0,0.38735,10.182713912027248,446.8108193616263
4,28.5,12.0,140.0,0.38735,17.761257602671837,687.8623353384033
\end{filecontents*}
\DTLloaddb[noheader=false]{obs31}{res/s9_D2/results/obs_31.csv}
\begin{filecontents*}{res/s9_D2/results/obs_32.csv}
idx,t,xo,yo,ro,xv,yv
1,0.0,12.0,145.0,0.38735,6.0,0.0
2,9.5,12.0,145.0,0.38735,11.178666434744933,194.6708036977185
3,19.5,12.0,145.0,0.38735,10.182713912027248,446.8108193616263
4,28.5,12.0,145.0,0.38735,17.761257602671837,687.8623353384033
\end{filecontents*}
\DTLloaddb[noheader=false]{obs32}{res/s9_D2/results/obs_32.csv}
\begin{filecontents*}{res/s9_D2/results/obs_33.csv}
idx,t,xo,yo,ro,xv,yv
1,0.0,12.0,150.0,0.38735,6.0,0.0
2,9.5,12.0,150.0,0.38735,11.178666434744933,194.6708036977185
3,19.5,12.0,150.0,0.38735,10.182713912027248,446.8108193616263
4,28.5,12.0,150.0,0.38735,17.761257602671837,687.8623353384033
\end{filecontents*}
\DTLloaddb[noheader=false]{obs33}{res/s9_D2/results/obs_33.csv}
\begin{filecontents*}{res/s9_D2/results/obs_34.csv}
idx,t,xo,yo,ro,xv,yv
1,0.0,12.0,155.0,0.38735,6.0,0.0
2,9.5,12.0,155.0,0.38735,11.178666434744933,194.6708036977185
3,19.5,12.0,155.0,0.38735,10.182713912027248,446.8108193616263
4,28.5,12.0,155.0,0.38735,17.761257602671837,687.8623353384033
\end{filecontents*}
\DTLloaddb[noheader=false]{obs34}{res/s9_D2/results/obs_34.csv}
\begin{filecontents*}{res/s9_D2/results/obs_35.csv}
idx,t,xo,yo,ro,xv,yv
1,0.0,12.0,160.0,0.38735,6.0,0.0
2,9.5,12.0,160.0,0.38735,11.178666434744933,194.6708036977185
3,19.5,12.0,160.0,0.38735,10.182713912027248,446.8108193616263
4,28.5,12.0,160.0,0.38735,17.761257602671837,687.8623353384033
\end{filecontents*}
\DTLloaddb[noheader=false]{obs35}{res/s9_D2/results/obs_35.csv}
\begin{filecontents*}{res/s9_D2/results/obs_36.csv}
idx,t,xo,yo,ro,xv,yv
1,0.0,12.0,165.0,0.38735,6.0,0.0
2,9.5,12.0,165.0,0.38735,11.178666434744933,194.6708036977185
3,19.5,12.0,165.0,0.38735,10.182713912027248,446.8108193616263
4,28.5,12.0,165.0,0.38735,17.761257602671837,687.8623353384033
\end{filecontents*}
\DTLloaddb[noheader=false]{obs36}{res/s9_D2/results/obs_36.csv}
\begin{filecontents*}{res/s9_D2/results/obs_37.csv}
idx,t,xo,yo,ro,xv,yv
1,0.0,12.0,170.0,0.38735,6.0,0.0
2,9.5,12.0,170.0,0.38735,11.178666434744933,194.6708036977185
3,19.5,12.0,170.0,0.38735,10.182713912027248,446.8108193616263
4,28.5,12.0,170.0,0.38735,17.761257602671837,687.8623353384033
\end{filecontents*}
\DTLloaddb[noheader=false]{obs37}{res/s9_D2/results/obs_37.csv}
\begin{filecontents*}{res/s9_D2/results/obs_38.csv}
idx,t,xo,yo,ro,xv,yv
1,0.0,12.0,175.0,0.38735,6.0,0.0
2,9.5,12.0,175.0,0.38735,11.178666434744933,194.6708036977185
3,19.5,12.0,175.0,0.38735,10.182713912027248,446.8108193616263
4,28.5,12.0,175.0,0.38735,17.761257602671837,687.8623353384033
\end{filecontents*}
\DTLloaddb[noheader=false]{obs38}{res/s9_D2/results/obs_38.csv}

\if\slides1
\newcommand{\hc}{3cm}
\newcommand{\wc}{7cm}

\newcommand{\hd}{5.5cm}
\newcommand{\wd}{6cm}
\else
\newcommand{\hc}{3cm}
\newcommand{\wc}{7cm}

\newcommand{\hd}{20cm}

\fi

\begin{figure}
\begin{center}
\begin{tikzpicture}
\pgfplotsset{
    height = \hd,
    width = 5cm,
    ymin=0, ymax=720,
    xmin=-75, xmax=100,
    scale only axis=true,
    axis equal image, 
    grid=major,
    ylabel={$y\;(\si{m})$},
    xlabel={$x\;(\si{m})$},
    cycle list name=my markers,
}
\coordinate (top) at (rel axis cs:0,1);
\begin{axis}[]

\draw[white!85!blue, fill] (-100,0) rectangle (0, 750);
\draw[white!85!blue, fill] (24,0) rectangle (100, 750);

\draw[->,black,very thick]  (-25,620) to (0,620);
\node [left] at (-25,620) {$x_{min}$};

\draw[->,black,very thick]  (50,620) to (24,620);
\node [right] at (50,620) {$x_{max}$};

\DTLforeach*{case9}{\xg=xRef, \yg=yRef, \goalTol=goalTol}{
\draw[white!50!green, fill] (\xg,\yg) circle (14);
\draw[->,black,very thick] (\xg+20,\yg) to (\xg,\yg);
\node [right] at (\xg+20,\yg) {Goal};

\addplot +[mark=none] table [x=x, y=y] {\move};
\addplot +[mark=none] table [x=x, y=y] {\static};
                                                }

\pgfplotsextra{
\DTLforeach*{obs1}{\x=xo, \y=yo, \r=ro, \t=t}{
\draw[thick, fill=red] (\x,\y) circle (\r);
}

\DTLforeach*{obs2}{\x=xo, \y=yo, \r=ro, \t=t,\clr=clr}{
\draw[thick, fill=pink!\clr!red] (\x,\y) circle (\r);
}
\DTLforeach*{obs3}{\x=xo, \y=yo, \r=ro, \t=t}{\draw[thick, fill=red] (\x,\y) circle (\r);}
\DTLforeach*{obs4}{\x=xo, \y=yo, \r=ro, \t=t}{\draw[thick, fill=red] (\x,\y) circle (\r);}
\DTLforeach*{obs5}{\x=xo, \y=yo, \r=ro, \t=t}{\draw[thick, fill=red] (\x,\y) circle (\r);}
\DTLforeach*{obs6}{\x=xo, \y=yo, \r=ro, \t=t}{\draw[thick, fill=red] (\x,\y) circle (\r);}
\DTLforeach*{obs7}{\x=xo, \y=yo, \r=ro, \t=t}{\draw[thick, fill=red] (\x,\y) circle (\r);}
\DTLforeach*{obs8}{\x=xo, \y=yo, \r=ro, \t=t}{\draw[thick, fill=red] (\x,\y) circle (\r);}
\DTLforeach*{obs9}{\x=xo, \y=yo, \r=ro, \t=t}{\draw[thick, fill=red] (\x,\y) circle (\r);}
\DTLforeach*{obs10}{\x=xo, \y=yo, \r=ro, \t=t}{\draw[thick, fill=red] (\x,\y) circle (\r);}
\DTLforeach*{obs11}{\x=xo, \y=yo, \r=ro, \t=t}{\draw[thick, fill=red] (\x,\y) circle (\r);}
\DTLforeach*{obs12}{\x=xo, \y=yo, \r=ro, \t=t}{\draw[thick, fill=red] (\x,\y) circle (\r);}
\DTLforeach*{obs13}{\x=xo, \y=yo, \r=ro, \t=t}{\draw[thick, fill=red] (\x,\y) circle (\r);}
\DTLforeach*{obs14}{\x=xo, \y=yo, \r=ro, \t=t}{\draw[thick, fill=red] (\x,\y) circle (\r);}
\DTLforeach*{obs15}{\x=xo, \y=yo, \r=ro, \t=t}{\draw[thick, fill=red] (\x,\y) circle (\r);}
\DTLforeach*{obs16}{\x=xo, \y=yo, \r=ro, \t=t}{\draw[thick, fill=red] (\x,\y) circle (\r);}
\DTLforeach*{obs17}{\x=xo, \y=yo, \r=ro, \t=t}{\draw[thick, fill=red] (\x,\y) circle (\r);}
\DTLforeach*{obs18}{\x=xo, \y=yo, \r=ro, \t=t}{\draw[thick, fill=red] (\x,\y) circle (\r);}
\DTLforeach*{obs19}{\x=xo, \y=yo, \r=ro, \t=t}{\draw[thick, fill=red] (\x,\y) circle (\r);}
\DTLforeach*{obs20}{\x=xo, \y=yo, \r=ro, \t=t}{\draw[thick, fill=red] (\x,\y) circle (\r);}
\DTLforeach*{obs21}{\x=xo, \y=yo, \r=ro, \t=t}{\draw[thick, fill=red] (\x,\y) circle (\r);}
\DTLforeach*{obs22}{\x=xo, \y=yo, \r=ro, \t=t}{\draw[thick, fill=red] (\x,\y) circle (\r);}
\DTLforeach*{obs23}{\x=xo, \y=yo, \r=ro, \t=t}{\draw[thick, fill=red] (\x,\y) circle (\r);}
\DTLforeach*{obs24}{\x=xo, \y=yo, \r=ro, \t=t}{\draw[thick, fill=red] (\x,\y) circle (\r);}
\DTLforeach*{obs25}{\x=xo, \y=yo, \r=ro, \t=t}{\draw[thick, fill=red] (\x,\y) circle (\r);}
\DTLforeach*{obs26}{\x=xo, \y=yo, \r=ro, \t=t}{\draw[thick, fill=red] (\x,\y) circle (\r);}
\DTLforeach*{obs27}{\x=xo, \y=yo, \r=ro, \t=t}{\draw[thick, fill=red] (\x,\y) circle (\r);}
\DTLforeach*{obs28}{\x=xo, \y=yo, \r=ro, \t=t}{\draw[thick, fill=red] (\x,\y) circle (\r);}
\DTLforeach*{obs29}{\x=xo, \y=yo, \r=ro, \t=t}{\draw[thick, fill=red] (\x,\y) circle (\r);}
\DTLforeach*{obs30}{\x=xo, \y=yo, \r=ro, \t=t}{\draw[thick, fill=red] (\x,\y) circle (\r);}
\DTLforeach*{obs31}{\x=xo, \y=yo, \r=ro, \t=t}{\draw[thick, fill=red] (\x,\y) circle (\r);}
\DTLforeach*{obs32}{\x=xo, \y=yo, \r=ro, \t=t}{\draw[thick, fill=red] (\x,\y) circle (\r);}
\DTLforeach*{obs33}{\x=xo, \y=yo, \r=ro, \t=t}{\draw[thick, fill=red] (\x,\y) circle (\r);}
\DTLforeach*{obs34}{\x=xo, \y=yo, \r=ro, \t=t}{\draw[thick, fill=red] (\x,\y) circle (\r);}
\DTLforeach*{obs35}{\x=xo, \y=yo, \r=ro, \t=t}{\draw[thick, fill=red] (\x,\y) circle (\r);}
\DTLforeach*{obs36}{\x=xo, \y=yo, \r=ro, \t=t}{\draw[thick, fill=red] (\x,\y) circle (\r);}
\DTLforeach*{obs37}{\x=xo, \y=yo, \r=ro, \t=t}{\draw[thick, fill=red] (\x,\y) circle (\r);}
\DTLforeach*{obs38}{\x=xo, \y=yo, \r=ro, \t=t}{\draw[thick, fill=red] (\x,\y) circle (\r);}

\draw[->,black,very thick]  (50,75) to (12,105);
\draw[->,black,very thick]  (50,75) to (12,75);
\draw[->,black,very thick]  (50,75) to (12,55);
\draw[->,black,very thick]  (50,75) to (12,25);
\node [right] at (50,75) {Cones};

\DTLforeach*{obs2}{\x=xo, \y=yo, \r=ro, \t=t, \idx=idx}{
\draw[->,black,very thick]  (\x+10,\y+15) to (\x,\y);
\node [right] at (\x+10,\y+15) {$O_2 \mathrm{,}\; \t$\;$\si{s}$};
}

\draw[->,black,very thick]  (-25,281) to (6,281);
\node [left] at (-25,281) {$O_1$};

\DTLforeach*{vehF9}{\xv=xv, \yv=yv, \psiV=psi, \t=t}{
\draw[blue, fill,rotate around={\psiV*180/pi:(\xv,\yv)}] (\xv-1.5,\yv-1) rectangle (\xv+1.5, \yv+1);
\node [above right] at (\xv+5,\yv) {$V_C \mathrm{,}\; \t$\;$\si{s}$};
                                                 }

\DTLforeach*{vehT9}{\xv=xv, \yv=yv, \psiV=psi, \t=t}{
\draw[orange, fill,rotate around={\psiV*180/pi:(\xv,\yv)}] (\xv-1.5,\yv-1) rectangle (\xv+1.5, \yv+1);
\node [above left] at (\xv-3,\yv-2) {$V_D \mathrm{,}\;\t$\;$\si{s}$};
                                                 }
               } 

\end{axis}

\coordinate (bot) at (rel axis cs:1,0);

\path (top|-current bounding box.north)--
      coordinate(legendpos)
      (bot|-current bounding box.north);
\matrix[
	fill=white,
    matrix of nodes,
    anchor=south,
    draw,
    inner sep=0.2em,
    draw
  ]at([yshift=-28ex,xshift=-14ex]legendpos)
  {
    \ref{plots:static} & $VC$ &[5pt]\\
    \ref{plots:move} & $VD$ & [5pt] \\
   };
\end{tikzpicture}
\end{center}

\if\slides0
    \caption{Closed-loop comparison of \plannerC and \plannerD in \caseCnsp. \label{fig:moa}}
\fi
\end{figure}

\begin{figure}
\begin{center}
\begin{tikzpicture}
\pgfplotsset{
    height = 5cm,
    width = 5cm,
    ymin=430, ymax=454,
    xmin=0, xmax=24,
    scale only axis=true,
    axis equal image, 
    grid=major,
    ylabel={$y\;(\si{m})$},
    xlabel={$x\;(\si{m})$},
    cycle list name=my markers,
}
\coordinate (top) at (rel axis cs:0,1);
\begin{axis}[]

\addplot +[mark=none] table [x=x, y=y] {\move};
\addplot +[mark=none] table [x=x, y=y] {\static};

\pgfplotsextra{

\DTLforeach*{obs2}{\x=xo, \y=yo, \r=ro, \t=t}{
\draw[thick, fill=red] (\x,\y) circle (\r);
}

\DTLforeach*{obs2}{\x=xo, \y=yo, \r=ro, \t=t, \idx=idx}{
\node [fill=white] at (\x-2,\y-2) {$O_2 \mathrm{,}\; \t$\;s};
}

\DTLforeach*{vehF9}{\xv=xv, \yv=yv, \psiV=psi, \t=t}{
\draw[blue, fill,rotate around={\psiV*180/pi:(\xv,\yv)}] (\xv-1.5,\yv-1) rectangle (\xv+1.5, \yv+1);
\node [right,fill=white] at (\xv+2,\yv) {$V_C \mathrm{,}\; \t$\;s};
                                                 }

\DTLforeach*{vehT9}{\xv=xv, \yv=yv, \psiV=psi, \t=t}{
\draw[orange, fill,rotate around={\psiV*180/pi:(\xv,\yv)}] (\xv-1.5,\yv-1) rectangle (\xv+1.5, \yv+1);
\node [left,fill=white] at (\xv-2,\yv) {$V_D \mathrm{,}\;\t$\;s};
                                                 }
               } 

\end{axis}

\coordinate (bot) at (rel axis cs:1,0);

\path (top|-current bounding box.north)--
      coordinate(legendpos)
      (bot|-current bounding box.north);
\matrix[
	fill=white,
    matrix of nodes,
    anchor=south,
    draw,
    inner sep=0.2em,
    draw
  ]at([yshift=-4cm,xshift=-1.5cm]legendpos)
  {
    \ref{plots:static} & $VC$ &[5pt]\\
    \ref{plots:move} & $VD$ & [5pt] \\
   };
\end{tikzpicture}
\end{center}

\if\slides0
	\caption{Zoomed in on Fig. \ref{fig:moc} at $19.5\;\si{s}$ \label{fig:mob}}
\fi
\end{figure}

\begin{figure}
\begin{center}
\begin{tikzpicture}

\begin{groupplot}[group style={group name = myplot, group size = 1 by 4, horizontal sep = 0.4cm, vertical sep = 1cm}, height = \hc, width = \wc]

\coordinate (top) at (rel axis cs:0,1);
\nextgroupplot[
             xmin = 1,
			 ymin = 0,
             ymax = 0.75,
   		     ylabel={solve-time $(\si{s})$},
             ylabel style={align=center, text width=2cm},
             xlabel={Iterations},
    	grid=major,
             cycle list name=my markers
             ]
\addplot +[mark=none] table [x=evalNum, y=tSolve] {\optT}; \label{plots:move}
\addplot +[mark=none] table [x=evalNum, y=tSolve] {\optF}; \label{plots:static}
\draw[dashed, very thick, blue] (0,0.5) -- (60,0.5);
\node at (30,.6) [fill=none] {Real-time threshold $=0.5\;\si{s}$};

\nextgroupplot[
             xmin = 0,
             ylabel={$\delta_f(t)\;(\si{radian})$},
             ylabel style={align=center, text width=2cm},
    	grid=major,
    	xticklabels=\empty,
             cycle list name=my markers
             ]
\addplot +[mark=none] table [x=t, y=sa] {\move};
\addplot +[mark=none] table [x=t, y=sa] {\static};

\nextgroupplot[
             xmin = 0,
             ylabel={ $u_x\;(\si{\frac{m}{s}})$},
             ylabel style={align=center, text width=2cm},
    	grid=major,
    	xticklabels=\empty,
             cycle list name=my markers
             ]
\addplot +[mark=none] table [x=t, y=ux] {\move};
\addplot +[mark=none] table [x=t, y=ux] {\static};

\nextgroupplot[
             xmin = 0,
             ylabel={$a_x\;(\si{\frac{m}{s^2}})$},
   	ylabel style={align=center, text width=2cm},
   	xlabel={Time $(\si{s})$},
    	grid=major,
             cycle list name=my markers
             ]
\addplot +[mark=none] table [x=t, y=ax] {\move};
\addplot +[mark=none] table [x=t, y=ax] {\static};

\end{groupplot}  

\coordinate (bot) at (rel axis cs:1,0);

\path (top|-current bounding box.north)--
      coordinate(legendpos)
      (bot|-current bounding box.north);
\matrix[
	fill=white,
    matrix of nodes,
    anchor=south,
    draw,
    inner sep=0.2em,
    draw
  ]at([yshift=-48ex,xshift=5ex]legendpos)
  {
    \ref{plots:static} & $V_C$ &[5pt]\\
    \ref{plots:move} & $V_D$ & [5pt] \\
   };
\end{tikzpicture}
\end{center}
\if\slides0
	\caption{Closed-loop comparison of \plannerC and \plannerD in \caseCnsp. \label{fig:moc}}
\fi
\end{figure}

\subsection{Execution Horizon and Obstacle Speed Analysis within \caseC} \label{sec:rd}
Including a moving obstacle avoidance specification increases safety over a range of execution horizons and obstacle speeds. To shown this $V_C$ and $V_D$ are tested within \caseC for a range of execution horizons ($\mathbf{t_{ex}}=[0.01,0.0621, \dots, 1]\;\si{s}$) and obstacle velocities ($\mathbf{v_y}[2]=[0,-2.11,\dots,-20]\;\frac{\si{m}}{\si{s}}$). The data from this parameter sweep are shown in Fig. \ref{fig:texv}, where a plotted point indicates a successful simulation. For instance, when the execution horizon is $0.01\;\si{s}$ and the velocity of $0_2$ is $-2.11\;\frac{\si{m}}{\si{s}}$, both $V_C$ and $V_D$ attain the goal.

The data follow the expected trend: i.e., $V_D$ is safer than $V_C$, and the results are statistically significant ($p=2.2\times10^{-16}$), as shown by a Fisher Test, in Appendix \ref{appendixa}. $V_D$ accounts for the majority ($87.1\%$) of the successful trials, and $V_C$ accounts for the majority of ($60.0\%$) of the trials that failed.

\pgfplotstableread{res/sweep2/results/sweepAoff.csv}{\static}
\pgfplotstableread{res/sweep2/results/sweepAon.csv}{\move}
\pgfplotstableread{res/sweep2/results/means.csv}{\means}
\pgfplotstableread{res/sweep2/results/probs.csv}{\probs}

\begin{figure}
\centering
\begin{tikzpicture}[only marks]
\pgfplotsset{
    height = 5cm,
    width = 8cm,
    grid=major,
    cycle list name=my markers ee,
    xlabel={Execution horizon$\;(\si{s})$},
    ylabel={$V_{obs} \;(\si{\frac{m}{s}})$},
    y dir=reverse
}
\coordinate (top) at (rel axis cs:0,1);
\begin{axis}[]

   \addplot table [
        x expr={
            ifthenelse(
                \thisrow{goalReached},
                \thisrow{tex},
                NaN
            )
        },
        y=v
    ] {\move}; \label{plots:Da}

    \addplot table [
        x expr={
            ifthenelse(
                \thisrow{goalReached},
                \thisrow{tex},
                NaN
            )
        },
       y=v
    ] {\static}; \label{plots:Ca}
\end{axis}
\coordinate (bot) at (rel axis cs:1,0);
\path (top|-current bounding box.north)--
      coordinate(legendpos)
      (bot|-current bounding box.north);
\matrix[
	fill=white,
    matrix of nodes,
    anchor=south,
    draw,
    inner sep=0.2em,
    draw
  ]at([yshift=-1cm,xshift=0cm]legendpos)
  {
    \ref{plots:Ca} & $VC$ & [5pt]  \ref{plots:Da} & $VD$ & [5pt] \\
   };
\end{tikzpicture}
	\caption{Effect that both the execution horizon and obstacle speed have on the vehicle attaining the goal in \caseC for both \plannerC and \plannerDnsp.
 \label{fig:texv}}
\end{figure}

While making the execution horizon small creates a more reactive planner, which can more reliably avoid collisions with fast moving obstacles, it makes it more difficult to obtain the planning solutions in real-time. Fig. \ref{fig:rtf} depicts this issue, where the real-time-factor (RTF) and probability-of-safety (POS) are defined as follows
\begin{definition}{Real-time-factor (RTF):}
$\mathrm{RTF}=\frac{\text{solve-times}_{\text{max}}}{t_{ex}}$. To calculate ${\text{solve-times}_{\text{max}}}$, the maximum value in a vector of solve-times for each test case (i.e., obstacle speed and execution horizon) is averaged across obstacle speeds.
\end{definition}
\begin{definition}{Probability-of-safety (POS):}
The probability that the vehicle will attain the goal, which is calculated over the range of obstacle velocities for each execution horizon.
\end{definition}

For both \plannerC and \plannerDnsp, the RTF is very high at small execution horizons and drops for larger execution horizons, as shown in Fig. \ref{fig:rtf}. For \plannerCnsp, the POS is very low across the entire range of execution horizons. In contrast, \plannerDnsp's POS is higher for smaller execution horizons and lower for larger execution horizons. Additionally, when using \plannerD there are two cases where the RFT is less than $1$, namely when the execution horizon is either $0.687$ or $0.790\;\si{s}$. In these cases the POS is $0.45$ and $0.25$, respectively.

\pgfplotstableread{res/sweep2/results/sweepAoff.csv}{\static}
\pgfplotstableread{res/sweep2/results/sweepAon.csv}{\move}
\pgfplotstableread{res/sweep2/results/means.csv}{\means}
\pgfplotstableread{res/sweep2/results/probs.csv}{\probs}

\begin{figure}
\centering
\begin{tikzpicture}
     \begin{axis}[
        black,
        hide x axis,
        height = 5cm,
        width = 7cm,
        xmin = 0, xmax = 1,
        ylabel={Real-time-factor},
        grid=major,
        cycle list name=my markersA,
        ymode=log,
        log basis y={10},
        axis y line*=left,
        xlabel near ticks,
        ylabel near ticks,
     ]
         \addplot[color=black,thin, mark=square] table [x=tex, y=maxoff] {\means}; \label{plots:Ac}
         \addplot[color=black,thin, mark=otimes] table [x=tex, y=maxon] {\means}; \label{plots:Bc}
     \end{axis}

     \coordinate (top) at (rel axis cs:0,1);
     \begin{axis}[
        red,
        hide x axis,
        height = 5cm,
        width = 7cm,
        xmin = 0, xmax = 1,
        ymin = 0, ymax = 1,
        y dir=reverse,
        hide x axis,
        axis y line*=right,
        ylabel={Probability-of-safety},
        ylabel near ticks,
     ]
        \addplot[color=red,thin,mark=square] table [x=tex, y=pc] {\probs}; \label{plots:Cc}
        \addplot[color=red,thin,mark=otimes] table [x=tex, y=pd] {\probs}; \label{plots:Dc}
     \end{axis}
     \coordinate (bot) at (rel axis cs:1,0);

     \path (top|-current bounding box.north)--
           coordinate(legendpos)
           (bot|-current bounding box.north);
     \matrix[
     	fill=white,
         matrix of nodes,
         anchor=south,
         draw,
         inner sep=0.2em,
         draw
       ]at([yshift=-0.05cm,xshift=0cm]legendpos)
       {
        \ref{plots:Ac} & \plannerC RTF & [5pt] & \ref{plots:Cc} & $VC$ POS & [5pt] \\
        \ref{plots:Bc} & \plannerD RTF & [5pt] & \ref{plots:Dc} & $VD$ POS & [5pt] \\
        };

\begin{axis}[
   height = 5cm,
   width = 7cm,
   grid=major,
   xmin = 0, xmax = 1,
   xlabel={Execution horizon$ \;(\si{s})$},
   ymode=log,
   hide y axis,
   log basis y={10},
   xlabel near ticks,
   ylabel near ticks,
]
\end{axis}

\begin{axis}[
   height = 5cm,
   width = 7cm,
   xmin = 0, xmax = 1,
   ymin = 0, ymax = 1,
   y dir=reverse,
   hide x axis,
   axis y line*=right,
   ylabel near ticks,
]
\end{axis}

\end{tikzpicture}
	\caption{Effect of the execution horizon on both the maximum real-time factor (left axis) and the probability-of-safety (right axis) in \caseC for both \plannerC and \plannerDnsp. \label{fig:rtf}}
\end{figure}

\section{Discussion} \label{sec:discussion}
This paper develops four NMPC-based trajectory planners, each with a different set of specifications. Comparisons among these planners, within three different environments, illuminate the potential effects of several key planner specifications on UGV safety and performance. These comparisons provide the basis for this paper's contributions.

This work was motivated by the assumption that including the set of specifications \reqAnsp-\reqG into a planner will improve both performance and safety, compared with less comprehensive sets. The results presented in this paper support this assumption. In particular, the results show that including (a) minimum time-to-goal, (b) minimum control effort, and (c) moving obstacle avoidance specifications improves the closed-loop performance and safety for a UGV application.

Contrary to our expectations, adding several key planner specifications does not lead to larger solve-times. Specifically, the results show that adding (a) minimum time-to-goal, (b) minimum control effort, and (c) moving obstacle avoidance specifications does not lead to an increase in NLP solve-times.

In fact, adding a minimum time-to-goal specification actually reduces the solve-times within the simple, static, unstructured environment (see  the top left trace in Fig. \ref{fig:mt}). The minimum time-to-goal specification helps balance the sixth term in Eqn. \ref{eqn:cost_a}, which minimizes the area between the vehicle's position trajectory and a line that runs through the goal in the y-direction. To see this balancing effect, compare the position trajectories of $V_A$ to those of $V_B$ and $V_C$ in Fig. \ref{fig:mt}. The baseline planner, i.e., \plannerAnsp, more effectively minimizes the area mentioned above for $V_A$ than either \plannerB or \plannerC does for $V_B$ and $V_C$, respectively. $V_B$ and $V_C$ have a larger area because both \plannerB and \plannerC have, in addition to the sixth term in Eqn. \ref{eqn:cost_a}, a minimum time-to-goal specification. To reduce this area more effectively, $V_A$ aggressively decelerates over the entire test and operates at lower speeds; these lower speeds allow $V_A$ to return to the line that runs through the goal in the y-direction sooner than either $V_B$ or $V_C$. These differences between the vehicle's trajectories may have led to the differences in the planners' solve-times, where \plannerA has longer solve-times than either \plannerB or \plannerCnsp. Notice that at around $6\;\si{s}$, $V_A$'s steering angle $\delta_f(t)$ and longitudinal deceleration $a_x(t)$ are large, and \plannerAnsp's solve-time increases sharply. On the basis of such observations, this paper speculates that planning aggressive deceleration and steering trajectories at low speeds may be more computationally expensive than planning less aggressive deceleration and steering trajectories at high speeds.

The results presented in this paper show that \NLOpt can solve UGV OCPs in real-time, suggesting that \NLOpt can solve complex OCPs faster than \MATLAB \cite{liu2017combined,febbo2017moving}. Our latest UGV work \cite{febbo2017moving} has a less complex OCP than this work, while using the same computer and the same class of collocation methods\footnote{In both cases local-collocation methods are used; this work uses the trapezoidal method and our previous work uses Euler's backward method.} as this work. Thus, the OCP solve-times obtained in this paper and our previous work can be compared to help evaluate the ability of the respective software stacks to quickly solve complex OCPs. Our previous work uses \MATLAB in conjunction with the \IPOPT NLP solver to solve a UGV planning problem. To illustrate a shortcoming of this work, Table \ref{tab:solve}, in Appendix \ref{appendixa}, summarizes long solve-times obtained using this software stack and hard and soft constraints for obstacle avoidance to solve a single OCP in dynamic, unstructured environments with $3$ and $17$ obstacles. Similar research shows that solving real-time UGV planning problems using \MATLAB and \IPOPT is challenging \cite{liu2017combined} --- planning problems are solved up to $30$ times slower than real-time with a $2.90\mathrm{GHz}$ Intel Xenon processor and a $0.5\;\si{s}$ execution horizon. This paper shows that solving UGV OCPs, using a direct-collocation method implemented in \NLOpt \cite{febbo_2017} in conjunction with the \KNITRO NLP solver, makes real-time solutions feasible. Additionally, unreported tests in \caseAnsp-\caseC indicate that using \NLOpt in conjunction with the open-source \IPOPT NLP solver yields similar solve-times. Therefore, this step forward for real-time UGV planning can be attributed to the novel design specifications of \NLOpt and not the \KNITRO NLP solver.

As the number of obstacles increase, the environment becomes more challenging, because the number of obstacles directly affects the computational load \cite{chiangsafety}. In the formulation developed in this work, the NLP dimensions grow linearly as the number of obstacles increases. Thus, increasing the number of obstacles from $3$ (in \caseA and \caseBnsp) to $38$ (in \caseCnsp) enlarges the size of the NLP and the computational load. It is reasonable to assume that this increase is a major factor in the corresponding increase of solve-times. To see this increase in solve-times, compare Fig. \ref{fig:mt} and Fig. \ref{fig:mo} to Fig. \ref{fig:moc}. Increasing the number of obstacles may result in a loss of real-time solutions. However, several approaches may be taken to use the formulation presented in this paper in an environment with many obstacles. These approaches include: developing a strategy that considers a smaller number of obstacles at a time, grouping several small, closely packed obstacles as a single obstacle, increasing computational power, or some combination of these.

For a given UGV, as obstacle speed increases, the environment becomes more challenging, because the vehicle is put in an increasingly difficult situation. The data plotted in Fig. \ref{fig:texv} support this claim; even with a moving obstacle avoidance specification, it is not possible to reliably avoid the oncoming obstacle $0_2$ in \caseC when it is moving faster than $21.1\;\frac{\si{m}}{\si{s}}$.

A moving obstacle avoidance specification may be unnecessary if the planner is updating quickly and the obstacles are moving slowly. The data plotted in Fig. \ref{fig:texv} also support this claim. The data reveals that if the obstacle is moving directly toward the vehicle at a speed less than $2.11\;\frac{\si{m}}{\si{s}}$, and if the execution horizon is less than $0.375\;\si{s}$, then a planner without a moving obstacle avoidance specification will safely attain the goal. Removing the moving obstacle avoidance specification will also eliminate the need for an algorithm to predict the speed of the obstacles. This simplification may be appropriate for some industrial applications, where the obstacles are known to move slowly.

In addition to a moving obstacle avoidance specification, as obstacle speed increases, a small execution horizon becomes increasingly crucial for safety. The data plotted in Fig. \ref{fig:texv} supports this claim as well; over a range of obstacle speeds, planning with a smaller execution horizon makes it more likely that the vehicle attains the goal. It is therefore desirable to make the execution horizon as small as possible in order to create a more reactive and safer planner. Having a small execution horizon, however, makes it more difficult for the planner to obtain solutions in real-time.

In order to ensure that the planning solutions are obtained in real-time while maintaining safety, it may be necessary to operate the UGV within environments where the obstacles are traveling from low to moderate speeds. Particular sets of data plotted in Fig. \ref{fig:texv} support this claim as well. Specifically, when disregarding the cases where the obstacle is traveling faster than $16.8\;\frac{\si{m}}{\si{s}}$, the RTF decreases to $0.933$ and the POS increases to $0.889$. Similarly, when disregarding the data where the obstacle is traveling faster than $4.21\;\frac{\si{m}}{\si{s}}$, the RTF is further reduced to $0.920$ and the POS increases to $1$.

In addition to reducing the execution horizon, planning in a dangerous situation can increase the RTF. This results from the fact that planning in a dangerous situation can lead to less feasible or even infeasible NLP constraints, which make it more challenging or even impossible for the NLP solver to obtain a solution. The top trace of Fig. \ref{fig:moc} supports this claim; it shows that solve-times increase sharply just before $V_D$ avoids a collision with $0_2$. It is important to consider these situations in terms of solve-time; if the planner cannot obtain a trajectory within the real-time limit, then the vehicle will not have a trajectory to follow and the situation status will go from dangerous to disastrous.
\section{Conclusion} \label{sec:conclusion}
This paper incorporates planner specifications \reqAnsp-\reqG (listed in  Table \ref{tab:cap}) into an NMPC-based trajectory planner for a UGV. UGV safety and performance is tested within four simulation-based comparisons. The results show that
\begin{itemize}
\item planners with less comprehensive sets of specifications than \reqAnsp-\reqG reduce UGV safety and performance,
\item if the planner is updating quickly, then a slowly moving obstacle can be safely avoided without a moving obstacle avoidance specification,
\item to avoid faster obstacles, both the moving obstacle avoidance and small execution horizon specifications are necessary,
\item a small execution horizon improves safety, but decreases the feasibility of obtaining trajectories in real-time, and
\item planning in an environment with more obstacles increases OCP solve-times.
\end{itemize}
Contrary to our expectations, our results show that adding the minimum-time-to-goal, minimum control effort, and moving obstacle avoidance specifications does not lead to larger solve-times. In fact, adding a minimum-time-to-goal specification actually reduces planning solve-times in the simple, static, unstructured environment. For our final research objective, the first three comparisons show that \NLOpt solves the OCP formulations, with a minimum-time-to-goal specification, in real-time, i.e., the solve-times are all less than the chosen execution horizon of $0.5\;\si{s}$. In contrast, previous work \cite{febbo2017moving,liu2017combined} shows that \MATLAB cannot solve OCP formulations that have a similar level of complexity in real-time. Therefore, \NLOpt is found to be a suitable tool for quickly solving complex OCPs. While this work tailors the NMPC-based trajectory planner for a UGV application, a variety of automated vehicle systems, e.g., UAVs and spacecraft, can also make use of the approach detailed here.
\bibliography{refs}
\if\thesis0
  \begin{appendices}
  \section{Supplementary Data} \label{appendixa}
  \else
    \appendix{Supplementary Material for Chapter \ref{MO2}}
\fi


\begin{figure}[ht]
\begin{center}
\captionof{table}{Long \MATLAB solve-times} \label{tab:solve}
\begin{tabular}{l| l l }
\hline
  &  \multicolumn{2}{c}{Solve-times} \\
\hline
Constraints & $3$ Obstacles &  $17$ Obstacles \\
\hline
Hard constraints & $44.4\;\si{s}$ & $193\;\si{s}$ \\
Soft constraints & $110\;\si{s}$ & $2.19\times10^{3}\;\si{s}$ \\
\hline
\end{tabular}
\end{center}
\end{figure}

\begin{figure}[ht]
\begin{center}
\captionof{table}{Vehicle Parameters} \label{tab:veh}
\begin{tabular}{ l l l}
\hline
Variable & Value & Units\\
\hline
$M_t$ & 2689 & \si{kg} \\
$I_{zz}$ & 4110 & \si{kg-m^2} \\
$L_f$,$L_r$ & $1.58$,$1.72$ & \si{m} \\
$K_{z_{x}}$,$K_{z_{yr}}$,$K_{z_{yf}}$ & $806$,$1076$,$675$ & $\frac{\si{N}}{\frac{\si{m}}{\si{s^2}}}$ \\
$F_{z_{min}}$ & $1000$ & \si{N} \\
$a$,$b$ & $1300$,$100$ & - \\
 $\psi_{\min}$,$\psi_{\max}$ & $[-2\pi,2\pi]$ & \si{\degree} \\
 $\delta_{f, \min}$,$\delta_{f, \max}$ & $[-30,30]$ & \si{\degree} \\
 $\gamma_{f, \min}$,$\gamma_{f, \max}$ & $[-5,5]$ &  $\frac{\si{\degree}}{\si{s}}$ \\
 $J_{x, \min}$,$J_{x, \max}$&$[-5,5]$ & $\frac{\si{m}}{\si{s^3}}$\\
$U_{\min}$,$U_{\max}$&$[0.01,29]$ & $\frac{\si{m}}{\si{s}}$\\
\end{tabular}
\end{center}
\end{figure}

\begin{figure*}
 \begin{minipage}{\textwidth}
\begin{center}
\captionof{table}{Simulation Parameters for \plannerA} \label{tab:parama}
\begin{tabular}{ l l l}
\hline
Variable or Conditions & Value and Units\\
\hline
$t_{ex}$,$N$,$L_\text{range}$,$\kappa$ & $0.5\;\si{s}$,$10$,$50.0\;\si{m}$,$5.0\;\si{m}$ \\
$sm_1$,$\;sm_2$,$\;sm$ & $2.5 \;\mathrm{m}$,$4 \;\mathrm{m}$,$2 \;\mathrm{m}$\\
$\mathbf{X0}$ &  $[200\;\si{m}, 0\;\si{m}, 0, 0, 1.57\si{rad}, 0, 17\;\frac{\si{m}}{\si{s}}, 0]$ \\
$\mathbf{X0}_{tol}$ & $[0.5\;\si{m},0.5\;\si{m},0.5,0.005,0.5,0.25,0.5\;\frac{\si{m}}{\si{s}},0.5]$ \\
$\mathbf{XF}_{tol}$ & $[5.0\;\si{m},5.0\;\si{m},NaN,NaN,NaN,\dots$ \\
&                             $ NaN,NaN,NaN]$  \\
$w_{ic}$,$w_{x0}$,$w_{y0}$,$w_{v0}$,$w_{r0}$,$w_{\psi0}$,$w_{sa0}$,$w_{ux0}$,$w_{ax0}$,$w_{xf}$ & $100$,$1$,$1$,$10$,$10$,$10$,$2$,$0.1$,$0.1$,$100$ \\
$w_{g}$,$w_{t}$,$w_{haf}$,$w_{F_z}$,$w_{ce}$,$w_{sa}$,$w_{sr}$,$w_{ax}$,$w_{jx}$ & $10$,$0$,$1$,$0.5$,$0$,$0.1$,$1$,$0.1$,$0.01$\\
moving obstacle avoidance constraint in Eqn. \ref{eqn:obs} & $false$ \\
\end{tabular}
\end{center}
\end{minipage}
\end{figure*}



\begin{figure*}
\begin{minipage}{\textwidth}
\begin{center}
\captionof{table}{Environment for \caseA} \label{tab:enva}
\begin{tabular}{ l l l l}
\hline
Variable & Description &  Value & Unit\\
\hline
$\mathbf{a_{obs}}$ & array of the obstacles semi-major axes  & $[5,4,2]$ & $\si{m}$\\
$\mathbf{b_{obs}}$ & array of the obstacles semi-minor axes & $[5,4,2]$ & $\si{m}$\\
$\mathbf{x0_{obs}}$ & array of the obstacles initial $x$ positions  & $[205,180,200]$& $\si{m}$ \\
$\mathbf{y0_{obs}}$ & array of the obstacles initial $y$ positions & $[57,75,63]$& $\si{m}$ \\
$\mathbf{v_x}$ & array of obstacles speeds in $x$ direction  & $[0,0,0]$ & $\frac{\si{m}}{\si{s}}$\\
$\mathbf{v_y}$ & array of obstacles speeds in $y$ direction  & $[0,0,0]$ & $\frac{\si{m}}{\si{s}}$\\
$x_{g}$ & $x$ position of goal location  & $200$ & $\si{m}$\\
$y_{g}$ & $x$ position of goal location & $125$ & $\si{m}$\\
$\sigma$ & tolerance on goal location &  $15$ & $\si{m}$\\
$\psi_{g}$ & desired orientation at goal  & $\frac{\pi}{2}$& $\si{rad}$ \\
\end{tabular}
\end{center}
\end{minipage}
\end{figure*}


\begin{figure*}
\begin{minipage}{\textwidth}
\begin{center}
\captionof{table}{Environment for \caseC \\
see Table. \ref{tab:enva} for Variable Descriptions} \label{tab:envc}
\begin{tabular}{ l l l}
\hline
Variable &  Value & Unit\\
\hline
$\mathbf{a_{obs}}$  & $[6,6,0.387,0.387,0.387,0.387,0.387,0.387,0.387,0.387,0.387,0.387,0.387,0.387,\dots $ & $\si{m}$  \\
                    & $\dots 0.387,0.387,0.387,0.387,0.387,0.387,0.387,0.387,0.387,0.387,0.387,0.387,\dots$ & \\
                    & $\dots 0.387,0.387,0.387,0.387,0.387,0.387,0.387,0.387,0.387,0.387,0.387,0.387]$ & \\
$\mathbf{b_{obs}}$ & same as $\mathbf{a_{obs}}$ & $\si{m}$ \\
$\mathbf{x0_{obs}}$   & $[6,18,12,12,12,12,12,12,12,12,12,12,12,12,12,12,12,12,12,12,12,12,12,\dots$ &  $\si{m}$ \\
                   & $\dots 12,12,12,12,12,12,12,12,12,12,12,12,12,12,12]$ & \\
$\mathbf{y0_{obs}}$  & $ [281,650,0,5,10,15,20,25,30,35,40,45,50,55,60,65,70,75,80,85,90,\dots$ & $\si{m}$ \\
                     & $\dots 95,100,105,110,115,120,125,130,135,140,145,150,155,160,165,170,175]$ & \\
$\mathbf{v_x}$   & $[0,0,0,0,0,0,0,0,0,0,0,0,0,0,0,0,0,0,0,0,0,0,0,0,0,0,0,0,0,0,0,0,0,0,0,0,0,0]$ & $\frac{\si{m}}{\si{s}}$\\
$\mathbf{v_y}$   & $[0,-10,0,0,0,0,0,0,0,0,0,0,0,0,0,0,0,0,0,0,0,0,0,0,0,0,0,0,0,0,0,0,0,0,0,0,0,0]$ & $\frac{\si{m}}{\si{s}}$\\
$x_{g}$ & $18$ & $\si{m}$\\
$y_{g}$  & $700$ & $\si{m}$\\
$\sigma$  &  $25$ & $\si{m}$\\
$\psi_{g}$  & $\frac{\pi}{2}$ & $\si{rad}$ \\
\end{tabular}
\end{center}
\end{minipage}
\end{figure*}

\begin{figure}[!htb]
\begin{center}
\captionof{table}{Control Effort} \label{tab:cer}
\begin{tabular}{l l l l}
\hline
Effort Term       & \plannerB   & \plannerC & Decrease \\
\hline
Steering Angle    & $0.000586$ & $0.000422$    & $28.0$ $\%$ \\
Steering Rate     & $0.00129$ & $0.000922$ & $28.5$ $\%$ \\
Longitudinal Jerk & $0.530$  & $0.420$ & $20.8$ $\%$ \\
Total             & $0.532$ & $0.421$ & $20.9$ $\%$ \\
\end{tabular}
\end{center}
\end{figure}

\begin{figure}
\begin{center}
\captionof{table}{Fisher's exact test for attaining the goal for \plannerC and \plannerD in \caseC ($p=2.2\times10^{-16}$)} \label{tab:fishera}
\begin{tabular}{l l l l}
\hline
& Fail & Pass & Total\\
\hline
\plannerC & $378$ ($60.0\;\%$) & $22$ ($12.9\;\%$) & $400$ \\
\plannerD & $252$ ($40.0\;\%$) & $148$ ($87.1\;\%$) & $400$ \\
Total & $630$ ($\;100\%$) & $170$ ($100\;\%$) & $800$ \\
\end{tabular}
\end{center}
\end{figure}

\if\thesis0
\end{appendices}
\fi
\end{document}

%% file: figs/threeDOF.tex
\begin{figure}
\begin{center}

\begin{tikzpicture}

\pgfmathsetmacro{\y}{4.5}
\pgfmathsetmacro{\x}{2.5}
\pgfmathsetmacro{\Angle}{atan2(\x,\y)}
\pgfmathsetmacro{\AngleR}{30}
\pgfmathsetmacro{\AngleF}{45}
\pgfmathsetmacro{\AngleDelta}{25}
\pgfmathsetmacro{\COMradius}{0.15}

\coordinate (Origin) at (0,0);
\draw [-latex,thick] (Origin)--++(0,\x+1.5) coordinate (yaxis);
\draw [-latex,thick] (Origin)--++(\y+1.5,0) coordinate (xaxis);
\draw [dashed] (Origin)--++(\Angle:6.5cm) coordinate (AngleEnd);
\draw [dotted,thick] (\y,0) node [below] {$x$} --++(0,\x) coordinate (Fyf);
\draw [dotted,thick] (0,\x) node [left] {$y$} --++(\y,0);
\draw pic["$\Psi$", draw=black, text=black, -latex, angle eccentricity=1.25, angle radius=0.8cm]
              {angle=xaxis--Origin--Fyf};

\coordinate (Fyr) at ($ (Origin) + (\Angle:2cm) $);
\node at (Fyr) [rotate=\Angle,draw,thick,rounded corners=2mm,minimum width=1cm, minimum height=0.4cm] {};
\draw [red,-latex,thick] (Fyr)--++(\Angle+\AngleR:1.3cm) coordinate (RedArrowOne);
\draw pic["$\alpha_r$", draw=black, text=red, -latex, angle eccentricity=1.45, angle radius=0.8cm]
              {angle=Fyf--Fyr--RedArrowOne};
\draw [latex-,thick,blue] (Fyr)--++(\Angle-90:0.5cm) node [rotate=\Angle,right] {$F_{yr}$};

\draw [thick] (Fyr)--(Fyf);

\node at (Fyf) [rotate=\Angle+\AngleDelta,draw,thick,rounded corners=2mm,minimum width=1cm, minimum height=0.4cm] {};
\draw [red,-latex,thick] (Fyf)--++(\Angle+\AngleF:1.3cm) coordinate (RedArrowTwo);
\draw [dashed] (Fyf)--++(\Angle+\AngleDelta:1.3cm) coordinate (DeltaAngleEnd);
\draw pic["$\delta_f$", draw=black, text=black, -latex, angle eccentricity=1.35, angle radius=0.8cm]
           {angle=AngleEnd--Fyf--DeltaAngleEnd};
\draw pic["$\alpha_f$", draw=black, text=red, -latex, angle eccentricity=1.45, angle radius=1cm]
      {angle=DeltaAngleEnd--Fyf--RedArrowTwo};
\draw [latex-,thick,blue] (Fyf)--++(\Angle-90:0.5cm) node [rotate=\Angle,right] {$F_{yf}$};

\coordinate (COM) at ($ (Origin) + (\Angle:3.7cm) $);
\begin{scope}[rotate=\Angle]
\fill [radius=\COMradius] (COM) -- ++(\COMradius,0) arc [start angle=0,end angle=90] -- ++(0,-2*\COMradius) arc [start angle=270, end angle=180];
\draw [thick,radius=\COMradius] (COM) circle;
\end{scope}
\draw [-latex,thick,green] (COM)--++(\Angle+90:0.5cm) node [left,rotate=\Angle] {$V$};
\draw [-latex,thick,green] (COM)--++(\Angle:0.8cm) node [below,rotate=\Angle] {$U$};
\draw[-latex,thick,green,rounded corners=2mm,minimum width=1cm, minimum height=0.4cm] let \p1=($(Fyf)-(Fyr)$),\n1={-180+atan2(\y1,\x1)},\n2={\n1+180} in
($($(COM)!5mm!00:(Fyr)$)+({cos(\n1+90)*1mm},{sin(\n1+90)*1mm})$) arc(\n1:\n2:5mm)
node[midway,below,green]{$\omega_z$};

\coordinate (LrLabel) at ($ (Fyr) +  (\Angle+90:1cm) $);
\coordinate (COMLabel) at ($ (COM) +  (\Angle+90:1cm) $);
\coordinate (LfLabel) at ($ (Fyf) +  (\Angle+90:1cm) $);
\draw [{Bar}{latex}-{latex}{Bar}] (LrLabel)--(COMLabel) node [midway,sloped,fill=white] {$L_r$};
\draw [{Bar}{latex}-{latex}{Bar}] (COMLabel)--(LfLabel) node [midway,sloped,fill=white] {$L_f$};

\end{tikzpicture}

\caption{$3$DoF dynamic vehicle model \cite{liu2016study}. } \label{fig:model}
\end{center}
\end{figure}